\documentclass[12pt,preprint,longnamesfirst]{aastex}

\slugcomment{Astrophysical Journal, Accepted}

\shorttitle{SN Nucleosynthesis with Neutrino Oscillation}
\shortauthors{Yoshida et al.}

\begin{document}


\title{Neutrino Oscillation Effects on Supernova Light Element Synthesis}


\author{Takashi Yoshida\altaffilmark{1,2}, Toshitaka Kajino\altaffilmark{3,4},
Hidekazu Yokomakura\altaffilmark{5}, Keiichi Kimura\altaffilmark{5}, \\
Akira Takamura\altaffilmark{6}, Dieter H. Hartmann\altaffilmark{7}}
\affil{$^1$Astronomical Institute, Graduate School of Science, 
Tohoku University, Aramaki, Aoba-ku, Sendai, Miyagi 980-8578, Japan}
\affil{$^2$National Astronomical Observatory of Japan, 2-21-1 Osawa, Mitaka, 
Tokyo 181-8588, Japan}
\email{takashi.yoshida@nao.ac.jp}
\affil{$^3$National Astronomical Observatory of Japan, and The Graduate 
University for Advanced Studies, 2-21-1 Osawa, Mitaka, Tokyo 181-8588, Japan}
\affil{$^4$Department of Astronomy, Graduate School of Science,
University of Tokyo, 7-3-1 Hongo, Bunkyo-ku, Tokyo 113-0033, Japan}
\affil{$^5$Department of Physics, Graduate School of Science,
Nagoya University, Furo-cho, Chikusa-ku, Nagoya, Aichi 464-8602, Japan}
\affil{$^6$Department of Mathematics, Toyota National College of Technology,
Eisei-cho 2-1, Toyota, Aichi 471-8525, Japan}
\affil{$^7$Department of Physics and Astronomy, Clemson University, Clemson,
SC 29634, USA}




\begin{abstract}
Neutrino oscillations affect light element synthesis through the $\nu$-process
in supernova explosions.
The $^7$Li and $^{11}$B yields produced in a supernova explosion of a 16.2
$M_\odot$ star model increase by factors of 1.9 and 1.3 in the case of
large mixing angle solution with normal mass hierarchy and 
$\sin^{2}2\theta_{13} \ga 2 \times 10^{-3}$ compared with those without
the oscillations.
In the case of inverted mass hierarchy or nonadiabatic 13-mixing resonance,
the increment of their yields is much smaller.
Neutrino oscillations raise the reaction rates of charged-current
$\nu$-process reactions in the region outside oxygen-rich layers.
The number ratio of $^7$Li/$^{11}$B could be a tracer of normal mass 
hierarchy and relatively large $\theta_{13}$, still satisfying 
$\sin^{2}2\theta_{13} \le 0.1$, through future precise 
observations in stars having strong supernova component.
\end{abstract}





\section{Introduction}

In the final evolutionary stage of massive stars, most region of the stars
except collapsing core explodes as supernova explosions.
The collapsing core releases its gravitational energy with gigantic
amount of neutrinos.
The emitted neutrinos interact with nuclei in the 
exploding material and new species of nuclei are produced; this synthetic
process is called the $\nu$-process \citep{de78,wh90}.
There are several species produced through the $\nu$-process.
For light elements, $^7$Li and $^{11}$B are mainly produced through the 
$\nu$-process (Woosley et al. 1990; Woosley \& Weaver 1995; 
Yoshida, Emori, \& Nakazawa 2000; Rauscher et al. 2002;
Yoshida et al. 2004; Yoshida, Kajino, \& Hartmann 2005).
Some $^{19}$F is also produced through the $\nu$-process 
\citep{wh90,ww95,rh02}.
For neutron-deficient heavy nuclei $^{138}$La and $^{180}$Ta are also 
produced through charged-current interactions with $\nu_e$
\citep{ga01,hk05}.
Neutrino-driven winds from proto-neutron stars are considered to be one
of the promoting sites for $r$-process heavy elements
\citep[e.g.,][]{ww94,tw94,ot00,tl04}.

Supernova explosion is one of the important sites for supplying $^7$Li and
$^{11}$B as well as Galactic cosmic rays, AGB stars, and novae during
Galactic chemical evolution (GCE) \citep[e.g.,][]{fo00}.
In previous studies, we showed that the amounts of $^7$Li and $^{11}$B 
strongly depend on the neutrino energy spectra and the total neutrino 
energy \citep{yt04,yk05}.
We also constrained the neutrino energy spectra from the gravitational
energy of a proto-neutron star and GCE models \citep{yk05}.
In these studies it has been assumed that the neutrino spectra do not
change in the supernova ejecta.

On the other hand, recent remarkable progress in neutrino experiments has
confirmed the phenomenon of neutrino oscillations \citep[e.g.,][]{mv04}.
The experiments on atmospheric neutrinos \citep[e.g.,][]{SK04},
solar neutrinos \citep[e.g.,][]{SNO04}, and reactor neutrinos
\citep[e.g,][]{ap03,KL05} constrained most of parameter values in 
neutrino oscillations, such as squared mass differences and the mixing angles.
Resultantly, the large mixing angle (LMA) solution turns out to be a
unique solution for the 12- and 23-mixings.
However, mass hierarchy between 1 and 3 mass eigenstates has not
been clarified \citep[e.g.,][]{SK04} and only upper limit of 
$\sin^{2}2\theta_{13}$ has been determined \citep{ap03}.

Supernova neutrino is another promoting target for neutrino experiments.
When SN 1987A occurred, Kamiokande group and IMB group found eleven and eight
events of neutrino detection \citep{hk87,IMB87}.
Owing to the development of neutrino experiments, much larger events of the
neutrino detection are expected when a supernova explosion occurs in 
neighboring galaxies.
In order to evaluate the neutrino flux and their energy dependence, neutrino
oscillations in supernova explosions have been investigated qualitatively
\citep{ds00} and quantitatively \citep{tw01}.
They showed that in the case of adiabatic resonance for 
$\sin^{2}2\theta_{13}$ (LMA-L in Takahashi et al. 2001) the transition 
probability from $\nu_e$ to $\nu_{\mu,\tau}$ changes from 0 to almost 1 
in the O/C layer in their supernova model.
Finally, the energy spectrum of $\nu_e$ changes to the one close to the 
$\nu_{\mu,\tau}$ spectrum emitted from neutrino sphere.
In the case of nonadiabatic resonance for $\sin^{2}2\theta_{13}$
(LMA-S in Takahashi et al. 2001), the change of neutrino spectra is smaller.
The effects from mass hierarchy and from the change of the density profile
due to the shock propagation were also investigated \citep{ts03a,ts03b}.

Since neutrino oscillations change neutrino spectra, it is expected
that the amounts of $^7$Li and $^{11}$B change by the effect of neutrino
oscillations \citep{yk06}.
During supernova explosions, neutral-current reactions such as
$^4$He($\nu,\nu'p)^3$H and $^4$He($\nu,\nu'n)^3$He are important
for $^7$Li and $^{11}$B production \citep{yt04}.
We note here that the total reaction rates of neutral-current reactions
do not change by the neutrino oscillations.
The energy spectrum summed up in all neutrinos and antineutrinos does not 
change by the oscillations.
On the other hand, the reaction rates of charged-current reactions
such as $^4$He($\nu_e,e^-p)^3$He and $^4$He($\bar{\nu}_e,e^+n)^3$H are 
expected to increase by neutrino oscillations.
As shown in \citet{ds00} and \citet{tw01}, the mean energies of
$\nu_e$ and $\bar{\nu}_e$ increase by the neutrino oscillations.
This increase will raise the efficiency of the $^7$Li and $^{11}$B production.
If we obtain some clear signals of neutrino oscillations in the abundances of
$^7$Li and $^{11}$B, we would constrain the parameter values of 
neutrino oscillations from observations of light elements.
This is a new procedure to constrain neutrino oscillation parameters
completely different from the detections of supernova neutrinos.

In the present study, we investigate light element synthesis in 
supernova explosions taking account of the change of the neutrino spectra
due to neutrino oscillations.
We also evaluate the dependence of the yields of $^7$Li and $^{11}$B
in the supernova ejecta on the mixing angle $\sin^{2}2\theta_{13}$
and mass hierarchy.

We set the luminosity and the energy spectrum of each flavor of neutrinos 
just emitted from a proto-neutron star in \S 2.
We also set the parameter values of neutrino oscillations from the
results of recent neutrino experiments.
We explain a supernova explosion model and a nuclear reaction network
for light element synthesis.
We mention the cross sections of charged-current reactions of the 
$\nu$-process to evaluate the reaction rates including neutrino oscillations.
In \S 3, we show the transition probabilities of neutrino flavors
with different values of $\sin^{2}2\theta_{13}$ and mass hierarchy.
We also discuss the effect of the oscillations on the reaction rates
of charged-current $\nu$-process reactions.
In \S 4, we show the calculated mass fraction distribution of $^7$Li and 
$^{11}$B taking account of neutrino oscillations.
Then, we show the dependence of the $^7$Li and $^{11}$B yields on 
$\sin^{2}2\theta_{13}$ and mass hierarchy.
We also show the dependence on the temperatures of $\nu_{\rm e}$ and
$\bar{\nu}_{\rm e}$ just emitted from proto-neutron star.
In \S 5, we discuss the neutrino oscillations with supernova 
shock propagation and show the change of the $^7$Li and $^{11}$B yields
by this effect.
We also discuss the $^7$Li and $^{11}$B yields related to observations of
stars which have traces of supernova explosions and supernova remnants.
Finally, we conclude our study in \S 6.

\section{Supernova Model and Parameters}

\subsection{Models of Supernova Neutrinos Emitted from a Proto-Neutron Star}

We use a model of supernova neutrinos just emitted from a proto-neutron
star based on the models in previous studies on the $\nu$-process
nucleosynthesis \citep{yt04,yk05,yk06}.
We here set up the energy spectra of three flavors of neutrinos when they 
just emitted from the proto-neutron star before they have been affected
by neutrino oscillations in passing through the envelope.
The neutrino luminosity is assumed to decrease exponentially with the
decay time of $\tau_\nu = 3$ s \citep[after][]{wh90}.
The total neutrino energy $E_\nu$ is set to be $3.0 \times 10^{53}$ ergs
which is almost equal to the gravitational binding energy of a 1.4 $M_\odot$
neutron star \citep[e.g.,][]{lp01}.
The neutrino luminosity is equally partitioned for each flavor of neutrinos.
The neutrino energy spectra are assumed to obey Fermi-Dirac (FD) 
distributions with zero-chemical potentials, similarly to previous studies 
on the $\nu$-process.
The influence of nonzero chemical potentials has been discussed in
\citet{yk05}.
The temperature of $\nu_{\mu,\tau}$ and $\bar{\nu}_{\mu,\tau}$,
$T_{\nu_{\mu,\tau}}$, is set to be 6.0 MeV.
This neutrino temperature is adopted so that the production of $^{11}$B from
supernovae satisfies appropriate for GCE constraint on the light elements 
\citep{yk05}.
The temperatures of $\nu_e$ and $\bar{\nu}_e$,
$T_{\nu_e}$ and $T_{\bar{\nu}_e}$, are set to be
3.2 MeV and 5.0 MeV, with which we have investigated
light element synthesis and the $r$-process heavy element synthesis
\citep{yt04}.
We investigate detailed influences of neutrino oscillations using this 
set of neutrino temperatures.

We also use the temperature of 4.0 MeV for $\nu_e$ and $\bar{\nu}_e$, which 
has been adopted in \citet{ww95}, \citet{rh02}, and others.
It has been indicated that the cooling of proto-neutron stars makes the
temperatures of each flavor of neutrinos closer to each other 
\citep[e.g.,][]{kr03}.
Therefore, we consider the case of $T_{\nu_e}$ and $T_{\bar{\nu}_e}$
to be 4.0 MeV and 5.0 MeV, respectively, too.
For comparison, we set $T_{\nu_e}$ and $T_{\bar{\nu}_e}$ as
3.2 MeV and 4.0 MeV, corresponding to the case where only 
$T_{\bar{\nu}_e}$ is changed.

\subsection{Parameters for Neutrino Oscillations}

In the present study, we consider the change of the neutrino spectra
due to 3-flavor neutrino oscillations
\begin{equation}
i\hbar c\frac{d}{dx}
\left( \begin{array}{c}
\nu_e \\
\nu_\mu \\
\nu_\tau
\end{array} \right)
=\left\{ U \left( \begin{array}{ccc}
0 & 0 & 0 \\
0 & \frac{\Delta m^2_{21} c^4}{2 \varepsilon_\nu} & 0 \\
0 & 0 & \frac{\Delta m^2_{31} c^4}{2 \varepsilon_\nu}
\end{array} \right) U^\dag + \left( \begin{array}{ccc}
\pm \sqrt{2} G_F (\hbar c)^3 \frac{\rho Y_e}{m_{\rm u}} & 0 & 0 \\
0 & 0 & 0 \\
0 & 0 & 0
\end{array} \right) \right\}
\left( \begin{array}{c}
\nu_e \\
\nu_\mu \\
\nu_\tau
\end{array} \right) ,
\end{equation}
\begin{equation}
U = \left( \begin{array}{ccc}
c_{12}c_{13} & s_{12}c_{13} & s_{13} \\
-s_{12}c_{23}-c_{12}s_{23}s_{13} & c_{12}c_{23}-s_{12}s_{23}s_{13} &
s_{23}c_{13} \\
s_{12}s_{23}-c_{12}c_{23}s_{13} & -c_{12}s_{23}-s_{12}c_{23}s_{13} &
c_{23}c_{13}
\end{array} \right) ,
\end{equation}
where $\Delta m^2_{ij} = m^2_i - m^2_j$, $m_i$ is the mass of $i$-eigenstate
neutrinos, $\varepsilon_\nu$ is the neutrino
energy, $G_F$ is Fermi constant, $\rho$ is the density, $Y_e$ is 
electron fraction, $m_{\rm u}$ is the atomic mass unit, 
$s_{ij}=\sin \theta_{ij}$, and $c_{ij} = \cos \theta_{ij}$.
Positive and negative signs in the potential term
$\pm \sqrt{2} G_F(\hbar c)^3 \rho Y_{e}/m_{{\rm u}}$ correspond
to the cases of neutrinos and antineutrinos.
In this formulation the squared mass differences and the mixing angles
are parameters.
The values of these parameters except $\theta_{13}$ and the sign of 
$\Delta m^2_{31}$ have been precisely determined from recent
neutrino experiments by Super-Kamiokande \citep{SK04}, SNO \citep{SNO04}, 
and KamLAND \citep{KL05}.
We assume the mass differences and the mixing angles
as follows:
\begin{equation}
\Delta m^2_{21} = 7.9 \times 10^{-5} \quad {\rm eV^2} \quad {\rm and}
\quad \Delta m^2_{31} = \pm 2.4 \times 10^{-3} \quad {\rm eV^2}
\end{equation}
and
\begin{equation}
\sin^{2}2\theta_{12} = 0.816 \quad {\rm and} \quad 
\sin^{2}2\theta_{23} = 1.0.
\end{equation}
This parameter set corresponds to the LMA solution of neutrino oscillations.
The sign of $\Delta m^2_{31}$ has not been determined from recent neutrino
experiments.
Thus, we consider both normal mass hierarchy and inverted mass hierarchy.
The positive value and the negative value of $\Delta m^2_{31}$ correspond
to normal mass hierarchy, i.e., $m_1 < m_2 < m_3$, and inverted mass
hierarchy, i.e., $m_3 < m_1 < m_2$, respectively.
From CHOOZ experiment \citep{ap03} only upper limit of $\sin^{2}2\theta_{13}$
has been determined.
We investigate the influence of changing $\theta_{13}$ in the range of 
\begin{equation}
0 \le \sin^{2}2\theta_{13} \le 0.1.
\end{equation}

\subsection{Supernova Explosion Model}

We calculate light element nucleosynthesis using the same supernova
explosion model adopted in our previous studies \citep{yt04,yk05}.
The presupernova model is 14E1 model, which is a 16.2 $M_\odot$ star
just before the supernova explosion and corresponds to 
SN 1987A \citep{sn90}.
The shock propagation of the supernova is calculated using a piecewise
parabolic method code \citep{cw84,sn92}.
The explosion energy is set to be $1 \times 10^{51}$ ergs.
The mass cut is located at 1.61 $M_\odot$.
Note that it is not needed to calculate the structure inside the mass cut.
In the present study, we calculate neutrino oscillations with the density
structure of the presupernova model.
When the shock front is in inner high density region, there is no influence 
of the shock propagation to the neutrino oscillations.
This is because the oscillation amplitude is too small despite the 
presence of the shock wave in such a high density region.
Its effect will be discussed in detail in \S 5.

\subsection{Nuclear Reaction Network for the $\nu$-Process with
Neutrino Oscillations}

Our nuclear reaction network consists of 291 species of nuclei from
$n$, $p$, to Ge.
The included nuclear species and their associated nuclear reactions are 
the same as those in \citet{yt04}, except charged-current $\nu$-process 
reactions on $^4$He and $^{12}$C, which will be discussed later in this
subsection.
The rates of neutral-current $\nu$-process reactions and those of the
other charged-current reactions are adopted from the rates with the 
assumption of FD distribution of the neutrino spectra.
As mentioned in Introduction, the rates of the neutral-current reactions do 
not change by neutrino oscillations.

When we consider neutrino oscillations, the energy spectra of $\nu_e$
and $\bar{\nu}_e$ emitted from a proto-neutron star change
from FD distributions and charged-current $\nu$-process reaction rates
are not merely simple functions of neutrino temperatures.
In such a condition, we need to evaluate a neutrino flux as a function
of the neutrino energy.
The number flux of $i$-flavor neutrinos ($i=e, \mu, \tau$) emitted
from the proto-neutron star with the energy of $\varepsilon_\nu$ can
be written as
\begin{equation}
\frac{d\phi_{\nu_i}}{d\varepsilon_\nu} = \frac{L_{\nu_i}}{4 \pi r^2}
\frac{1}{F_3(\eta_{\nu_i})(kT_{\nu_i})^4}
\frac{\varepsilon_\nu^2}
{\exp(\frac{\varepsilon_\nu}{kT_{\nu_i}}-\eta_{\nu_i})+1}
\end{equation}
and
\begin{equation}
F_3(\eta_{\nu_i}) = \int^\infty_0
\frac{x^3 dx}{\exp(x-\eta_{\nu_i})+1},
\end{equation}
where $L_{\nu_i}$ is the luminosity of $i$-flavor neutrinos, $r$ is the 
radius, $k$ is the Boltzmann constant, $T_{\nu_i}$ is the neutrino 
temperature, and $\eta_{\nu_i}$ is the degeneracy factor \citep[e.g.,][]{by05}.
Here we define $P_{ij}(r;\varepsilon_\nu)$ as the oscillation
probability from $i$-flavor to $j$-flavor at the radius of $r$ and 
the energy of $\varepsilon_\nu$.
Taking the cross section of a $\nu$-process reaction as a function of
$\varepsilon_\nu$, $\sigma_{\nu_e}(\varepsilon_\nu)$,
we evaluate the reaction rate with neutrino oscillations 
$\lambda_{\nu_e}$ as
\begin{equation}
\lambda_{\nu_e} = \sum_{i = e,\mu,\tau}
\frac{L_{\nu_i}}{4 \pi r^2} 
\frac{1}{F_3(\eta_{\nu_i})(kT_{\nu_i})^4}
\int^\infty_0 \frac{\varepsilon_\nu^2 P_{ie}(r;\varepsilon_\nu) 
\sigma_{\nu_e}(\varepsilon_\nu)}
{\exp(\frac{\varepsilon_\nu}{kT_{\nu_i}}-\eta_{\nu_i})+1} d\varepsilon_\nu .
\end{equation}
In the present study, we consider the effect on neutrino oscillations
for charged-current reactions of $^4$He, 
$^4$He($\nu_e,e^-p)^3$He and 
$^4$He($\bar{\nu}_e,e^+n)^3$H, and the reactions of $^{12}$C,
$^{12}$C($\nu_e,e^-p)^{11}$C,
$^{12}$C($\nu_e,e^-\gamma)^{12}$N,
$^{12}$C($\bar{\nu}_e,e^+n)^{11}$B, and
$^{12}$C($\bar{\nu}_e,e^+\gamma)^{12}$B.
However, the detailed data of the cross sections of these reactions
as functions of $\varepsilon_\nu$ have not been reported and one can find only
energy averaged reaction rates as functions of neutrino temperature.
Therefore, we adopt analytical approximation to the cross sections with
respect to $\varepsilon_\nu$ as
\begin{equation}
\sigma_{\nu_e}(\varepsilon_\nu) = \left\{
\begin{array}{ll}
a (\varepsilon_\nu - \varepsilon_{th})^b & {\rm for} \quad
\varepsilon_\nu \ge \varepsilon_{th} , \\
0 & {\rm for} \quad 
\varepsilon_\nu < \varepsilon_{th} ,
\end{array}
\right.
\end{equation}
where $\varepsilon_{th}$ is the threshold energy of each reaction.
Coefficients are determined so as to fit the rate of the corresponding
reaction with the assumption of the FD distributions to the
values tabulated in the 1992 work by R. D. Hoffman \& S. E. Woosley, HW92
\footnotemark.
\footnotetext{http://www-phys.llnl.gov/Research/RRSN/nu\_csbr/neu\_rate.html.}
The coefficients and the thresholds are listed in Table 1.
When we evaluate the reaction rates from the cross sections with 
assumptions of the FD distributions, these values are in reasonable agreement 
with those of HW92 within $\pm 9$ \%.

\section{Neutrino Oscillations in Presupernovae}

In order to evaluate the reaction rates of the charged-current $\nu$-process
reactions with neutrino oscillations, we numerically solve the oscillations
of the neutrinos passing through a presupernova star in Runge-Kutta method
and based on analytical expression of \citet{kt02a,kt02b}.
We use the density profile of 14E1 model of Shigeyama \& Nomoto (1990).
The density profile is shown in Fig. 1 in Shigeyama \& Nomoto
(1990).

When neutrinos pass through the stellar interior, there are resonances of 
neutrino oscillations where the transition probabilities change largely.
The resonance density $\rho_{{\rm res}}$ is determined from the squared
mass difference and the neutrino energy.
The resonance density is written as
\begin{equation}
\rho_{{\rm res}} Y_e =
\frac{m_{\rm u} \Delta m^2_{ji} c^4 \cos 2\theta_{ij}}
{2\sqrt{2} G_F (\hbar c)^3 \varepsilon_\nu}
= 6.55 \times 10^6 
\left( \frac{\Delta m^2_{ji}}{1 \, {\rm eV}^2} \right)
\left( \frac{1 \, {\rm MeV}}{\varepsilon_\nu} \right) \cos 2\theta_{ij}
\quad {\rm g \, cm^{-3}}.
\end{equation}
Electron number density $n_e$ relates to the density and electron
fraction through $n_e = \rho Y_e/m_{\rm u}$.
The transition probabilities depend on the adiabaticity of the resonance
strongly.
The adiabaticity is estimated using the adiabaticity parameter $\gamma$.
The flip probability $P_f$, which means the probability that a neutrino
in one mass eigenstate changes to another mass eigenstate is written as
(e.g., Dighe \& Smirnov 2000)
\begin{equation}
P_f = \exp \left( -\frac{\pi}{2}\gamma \right)
\end{equation}
and 
\begin{equation}
\gamma = \frac{\Delta m_{ji}^2 c^3}{2 \hbar \varepsilon_\nu}
\frac{\sin^{2}2\theta_{ij}}{\cos2\theta_{ij}}
\left| \frac{1}{(1/n_e)(dn_e/dr)} \right| .
\end{equation}
When $\gamma \gg 1$, i.e., the flip probability $P_f$ is very small
($P_f \ll 1$), the resonance is adiabatic.
When the flip probability is close to unity, the resonance is nonadiabatic.

Figure 1 shows the transition probability of $\nu_e$ to $\nu_e$,
$P_{{\rm ee}}$, the sum of the transition probabilities of 
$\nu_{\mu} \rightarrow \nu_e$ and
$\nu_{\tau} \rightarrow \nu_e$, $P_{{\rm xe}}$, and the corresponding
transition probabilities for antineutrinos $P_{{\rm eeb}}$ and
$P_{{\rm xeb}}$ in the case of $\varepsilon_\nu = 50$ MeV and normal mass 
hierarchy.
There are resonances of $\Delta m^{2}_{31}$ and $\theta_{13}$, which is called
H-resonance, at $\rho_{\rm res} = 6.3 \times 10^2$ g cm$^{-3}$ and of
$\Delta m^{2}_{21}$ and $\theta_{12}$, which is called L-resonance, at
$\rho_{\rm res} = 8.9$ g cm$^{-3}$.
The H-resonance is located in innermost region of
the He/C layer ($\sim 3.8 M_\odot$) and  the L-resonance is 
in the He/C layer ($\sim 5.4 M_\odot$).
Both of H- and L-resonances appear for neutrinos and do not appear for
antineutrinos.
The adiabaticity of H-resonance depends on the value of
$\sin^{2}2\theta_{13}$.
L-resonance is always adiabatic in our parameter set.
We see in the mass coordinate range of $M_r \la 3.8 M_\odot$ that
the amplitude of neutrino oscillation is very small and that the flavor
exchange practically does not occur for both neutrinos and 
antineutrinos.

In the mass coordinate region of $M_r \ga 3.8 M_\odot$, the 
characteristics of the transition probabilities depend on mass hierarchy
and the adiabaticity of H-resonance.
In the case of $\sin^{2}2\theta_{13} = 1 \times 10^{-2}$ (see Fig. 1{\it a}), 
H-resonance appears for neutrinos and is adiabatic.
The transition from $\nu_{e}$ to $\nu_{\mu,\tau}$ occurs completely, 
i.e., $P_{{\rm ee}}$ becomes almost zero in the He/C layer.
At the same time, the transition probability from $\nu_{\mu}$ or $\nu_{\tau}$
to $\nu_e$ becomes large, i.e., $P_{{\rm xe}}$ is close to 1.
On the other hand, there is no resonance for antineutrinos.
The transition probabilities $P_{{\rm eeb}}$ and $P_{{\rm xeb}}$ gradually
change in the He/C layer, and about 30\% of antineutrinos change flavors.

In the case of $\sin^{2}2\theta_{13} = 1 \times 10^{-6}$ (see Fig. 1{\it b}), 
H-resonance is nonadiabatic.
So, complete change from $\nu_e$ to $\nu_{\mu,\tau}$ does not occur.
The transition probabilities for neutrinos change gradually as a function of
the mass coordinate similarly to antineutrinos.
Finally, about 70\% of neutrinos change flavors.
The transition probabilities for antineutrinos are the same as the case of 
$\sin^{2}2\theta_{13} = 1 \times 10^{-2}$.

Figure 2 shows the transition probabilities in the case of inverted mass
hierarchy.
In this case, H-resonance appears for antineutrinos.
When $\sin^{2}2\theta_{13}$ is equal to $1 \times 10^{-2}$ 
(see Fig. 2{\it a}), H-resonance is adiabatic and almost all 
$\bar{\nu}_e$ change to $\bar{\nu}_{\mu}$ and $\bar{\nu}_{\tau}$.
The transition probability $P_{{\rm xeb}}$ also becomes close to unity.
On the other hand, neutrinos change their flavors gradually
in the He-layer because of no appearance of H-resonance.
When $\sin^{2}2\theta_{13}$ is equal to $1 \times 10^{-6}$, H-resonance is 
nonadiabatic.
The change of the transition probabilities with increasing the mass
coordinate is the same as that in the case of normal mass
hierarchy and the nonadiabatic H-resonance.
In the case of nonadiabatic resonance, the flavor change occurs as if
there is no resonance.

The increase in charged-current $\nu$-process reaction rates strongly 
depends on the adiabaticity of H-resonance.
If H-resonance is adiabatic, the reaction rates of 
charged-current reactions for $\nu_e$ (normal mass hierarchy) and 
$\bar{\nu}_e$ (inverted mass hierarchy) become much larger than 
those without the oscillations in the He-layer.
If H-resonance is nonadiabatic, the increase in the rates of corresponding 
charged-current reactions would be much smaller.
Therefore, the adiabaticity of H-resonance affects the
final yields of $^7$Li and $^{11}$B.

We note that the shock propagation effect on neutrino oscillations
in supernova ejecta is not considered in this study.
We do not consider that the change of neutrino oscillations
due to the shock propagation would change the yields of $^7$Li and $^{11}$B.
When neutrinos pass through the O/Ne layer, the amplitude of neutrino
oscillations is very small because the density profile, where the density
is much larger than H-resonance density, does not affect
neutrino oscillations.
Thus, when the shock wave moves inside the O/C layer, the shock wave does
not affect the transition probabilities of neutrinos and antineutrinos
even if we consider the shock propagation.
After the shock wave arrives at the O/C layer, the shock wave will change
the transition probabilities.
We set the decay time of the neutrino flux to be 3 s which is to be compared
with the shock arrival time to the O/C layer about 5 s.
Because of this time lag, more than 80\% of neutrinos pass through the 
supernova ejecta before the shock arrives at the O/C layer.
We will discuss the influence of the shock wave on neutrino oscillations 
in the supernova in \S 5.

\section{Yields of $^7$Li and $^{11}$B}

\subsection{Mass Fraction Distributions of $^7$Li and $^{11}$B}

We show the change of the mass fractions of $^7$Li and $^{11}$B
due to neutrino oscillations.
Figure 3 shows the mass fraction distributions of $^7$Li and $^{11}$B.
Panels ({\it a}) and ({\it b}) correspond to the case of normal mass hierarchy,
and panels ({\it c}) and ({\it d}) correspond to inverted mass hierarchy.
In these figures we show the mass fraction distributions of $^7$Li
and its isobar $^7$Be separately.
The mass fraction distributions of $^{11}$B and $^{11}$C are also drawn
separately.

Let us first discuss the case of normal mass hierarchy and 
$\sin^{2}2\theta_{13} = 1 \times 10^{-2}$ 
(adiabatic H-resonance for neutrinos) shown by thick lines in Figs. 3$a$ 
and 3$b$.
We find the increase in the mass fractions of $^7$Li, $^7$Be, $^{11}$B, and
$^{11}$C compared with those without the neutrino oscillations in the He layer.
The mass fraction of $^7$Li with the neutrino oscillations is larger by
about a factor of 1.2 than that without the oscillations in most region of 
the He layer (see Fig. 3$a$).
In the range of $4.3 M_\odot \la M_r \la 4.6 M_\odot$, the increment
degree is larger. 
In the O-rich layers of $M_r \la 3.8 M_\odot$, we do not see any clear
differences due to neutrino oscillations in the mass fraction.
The transition probabilities to other neutrino flavors are very small 
because of high density in this region (see Fig. 1).
The obtained yield of $^7$Li is $1.79 \times 10^{-7} M_\odot$.
When we do not consider neutrino oscillations, the yield is 
$1.50 \times 10^{-7} M_\odot$.
The yield of $^7$Li increases by a factor of 1.19 owing to the neutrino 
oscillations.

The mass fraction of $^7$Be is larger by about a factor of 2.5 than that 
without the neutrino oscillations in the innermost region of the He layer, 
$3.8 M_\odot \la M_r \la 4.3 M_\odot$ (see Fig. 3$a$).
In the range of $4.3 M_\odot \la M_r \la 4.6 M_\odot$, where the $^7$Be mass 
fraction increases with the mass coordinate, the increase in the $^7$Be mass
fraction due to the neutrino oscillations is more than a factor of 3.
In the range of $M_r \ga 4.6 M_\odot$ it is a factor of $2.5 \sim 3.2$.
Finally, the obtained yield of $^7$Be is $2.66 \times 10^{-7} M_\odot$.
Without the neutrino oscillations, the yield of $^7$Be is
$8.62 \times 10^{-8} M_\odot$.
The yield of $^7$Be increases by a factor of 3.1 owing to the neutrino 
oscillations.
Thus, the total yield of $^7$Li, the sum of the yields of $^7$Li and $^7$Be, 
is $4.45 \times 10^{-7} M_\odot$ with the neutrino oscillations and 
$2.36 \times 10^{-7} M_\odot$ without the oscillations.
The $^7$Li yield with the neutrino oscillations is larger by a factor of
1.89 than that without the oscillations.

The mass fraction of $^{11}$B in the range of $M_r \ga 4.3 M_\odot$ of the
He layer is larger by about a factor of $1.2 \sim 1.3$ than that without
the oscillations (see Fig. 3$b$).
Inside the range of the He layer, it increases by about a factor of 2.1.
The increase in the mass fraction of $^{11}$C in the He layer is about 
a factor of $3.4 \sim 3.9$. 
This increase is larger than that of $^{11}$B and is close to the factor 
of $^7$Be.
The $^{11}$B and $^{11}$C mass fractions in the O-rich layers of
$M_r \la 3.8 M_\odot$ are not affected strongly by the neutrino oscillations.
The yields of $^{11}$B and $^{11}$C in this case are 
$7.02 \times 10^{-7} M_\odot$ and $9.16 \times 10^{-8} M_\odot$.
Without the neutrino oscillations, the corresponding yields are
$5.63 \times 10^{-7} M_\odot$ and $6.29 \times 10^{-8} M_\odot$.
Thus, the total $^{11}$B yield increases by a factor of 1.27.

We discuss the production process of $^7$Li and $^{11}$B in the He layer
and the effect of the neutrino oscillations on the $\nu$-process reactions.
In the He layer, $^7$Li and $^7$Be are produced through 
$^4$He($\nu,\nu'p)^3$H($\alpha,\gamma)^7$Li and
$^4$He($\nu,\nu'n)^3$He($\alpha,\gamma)^7$Be, respectively.
The corresponding charged-current $\nu$-process reactions are
$^4$He($\bar{\nu}_e,e^+n)^3$H($\alpha,\gamma)^7$Li and 
$^4$He($\nu_e,e^-p)^3$He($\alpha,\gamma)^7$Be.
Most of $^{11}$B is produced through $^7$Li($\alpha,\gamma)^{11}$B
and the contribution from $^{12}$C($\nu,\nu'p)^{11}$B is small.
The $^{11}$B in the O-rich layers is also produced from $^{12}$C.
The $^{11}$C is produced through $^{12}$C($\nu,\nu'n)^{11}$C.
The charged-current $\nu$-process reactions from $^{12}$C are
$^{12}$C($\bar{\nu}_e,e^+n)^{11}$B and
$^{12}$C($\nu_e,e^-p)^{11}$C.
Main production processes of these light elements are also written in 
\citet{yt04,yk05,yk06}.
In the case of normal mass hierarchy and adiabatic H-resonance for neutrinos,
$\nu_e$ and $\nu_{\mu,\tau}$ are completely exchanged owing to the 
adiabatic mixing, and therefore the rates of 
$^4$He($\nu_e,e^-p)^3$He and 
$^{12}$C($\nu_e,e^-p)^{11}$C increase.
Thus, the mass fractions of $^7$Be and $^{11}$C become larger.
On the other hand, there are no resonances for antineutrinos.
The transition probability between $\bar{\nu}_e$ and 
$\bar{\nu}_{\mu,\tau}$ is small so that the rates of 
$^4$He($\bar{\nu}_e,e^+n)^3$H and
$^{12}$C($\nu_e,e^-p)^{11}$C scarcely become larger, and therefore 
the increase in the mass fractions of $^7$Li and $^{11}$B is small.
The increase by a factor of 2 in the mass fraction of $^{11}$B is due to
the production by way of $^7$Be($n,p)^7$Li($\alpha,\gamma)^{11}$B.

Secondly, we consider the case of normal mass hierarchy and 
$\sin^{2}2\theta_{13} = 1 \times 10^{-6}$ (nonadiabatic H-resonance for 
neutrinos) shown by thin lines in Figs. 3$a$ and 3$b$.
The increment degree of $^7$Li mass fraction compared with the case without
the neutrino oscillations becomes gradually larger with increasing the
mass coordinate (see Fig. 3$a$).
The maximum increment degree is 1.3 at the outer edge of the He layer.
The mass fraction of $^7$Be is larger by a factor of 1.2 compared with the
case without the neutrino oscillations where the mass fraction has the maximum 
value.
The increment degree becomes larger with increasing the mass coordinate, 
but the mass fraction is much smaller there.
Thus, the obtained yields of $^7$Li and $^7$Be are 
$1.65 \times 10^{-7} M_\odot$ and $1.02 \times 10^{-7} M_\odot$.
The total yield of $^7$Li is $2.67 \times 10^{-7} M_\odot$; it is larger 
by a factor of only 1.13 than the case without the neutrino oscillations.

The increment degree of the $^{11}$B mass fraction due to the neutrino 
oscillations gradually increases with the mass coordinate in the He layer
(see Fig. 3$b$).
However, increment degree is 1.2 at the maximum at the outer edge of 
the He layer.
This is due to smaller transition probability from $\bar{\nu}_{\mu,\tau}$ to
$\bar{\nu}_e$ (see Fig. 1$b$).
The dependence of increasing $^{11}$C mass fraction on the
mass coordinate is similar to that of $^{11}$B.
The maximum increment degree is 2.9.
The obtained yields of $^{11}$B and $^{11}$C are 
$5.75 \times 10^{-7} M_\odot$ and $6.47 \times 10^{-8} M_\odot$ and, 
therefore, the total $^{11}$B yield is $6.40 \times 10^{-7} M_\odot$.
It is only slightly larger by a factor of 1.02 compared with the case without 
the oscillations.

Thirdly, we consider the case of inverted mass hierarchy and 
$\sin^{2}2\theta_{13} = 1 \times 10^{-2}$ (adiabatic H-resonance for 
antineutrinos) shown by thick lines in Figs. 3$c$ and 3$d$.
The mass fraction of $^7$Li is larger by a factor of $1.4 \sim 1.5$
in most region of the He layer.
For $^{11}$B, the mass fraction increases by a factor of $1.1 \sim 1.4$.
The yields of $^7$Li and $^{11}$B are $2.22 \times 10^{-7} M_\odot$ and
$7.05 \times 10^{-7} M_\odot$.
Compared with those without the oscillations, these yields are larger by
factors of 1.48 and 1.25.
Although adiabatic H-resonance appears for antineutrinos and, therefore, 
the flavor change between $\bar{\nu}_e$ and $\bar{\nu}_{\mu,\tau}$ occurs 
completely, the increment degree of the $^7$Li and $^{11}$B mass fractions are 
smaller than those of $^7$Be and $^{11}$C in the corresponding case of normal 
mass hierarchy.
This is because the difference between the mean energies for $\bar{\nu}_e$ 
and $\bar{\nu}_{\mu,\tau}$ is smaller than that for $\nu_e$
and $\nu_{\mu,\tau}$.
This small difference of the mean neutrino energies causes
smaller enhancement of $^7$Li and $^{11}$B even in the adiabatic mixing.

The mass fractions of $^7$Be and $^{11}$C are larger than those without
the neutrino oscillations in the range of $M_r \ga 4.8 M_\odot$ 
in the He layer.
On the other hand, they are smaller than those without the neutrino 
oscillations inside the region.
This is because neutrons produced through 
$^4$He($\bar{\nu}_e, e^+n)^3$H, of which rate is enhanced by the
neutrino oscillations, decompose $^3$He, $^7$Be, and $^{11}$C.
The yields of $^7$Be and $^{11}$C are $8.20 \times 10^{-8} M_\odot$ and
$6.35 \times 10^{-8} M_\odot$.
The yield ratios to the case without the neutrino oscillations are 0.95 and
1.01 for $^7$Be and $^{11}$C.
The total yields of $^7$Li and $^{11}$B are $3.04 \times 10^{-7} M_\odot$
and $7.69 \times 10^{-7} M_\odot$, and their increment factors are
1.29 and 1.23.

Finally, we show the case of inverted mass hierarchy and 
$\sin^{2}2\theta_{13} = 1 \times 10^{-6}$ (nonadiabatic H-resonance for 
antineutrinos) represented by thin lines in Figs. 3$c$ and 3$d$.
The mass fraction distributions of all four nuclear species are the
same as the corresponding distributions in the case of normal mass
hierarchy and nonadiabatic H-resonance.
We showed in \S3 that the exchange probabilities for 
$\nu_e (\bar{\nu}_e) \rightarrow 
\nu_e (\bar{\nu}_e)$ and 
$\nu_e (\bar{\nu}_e) \rightarrow 
\nu_{\mu,\tau} (\bar{\nu}_{\mu,\tau})$ are identical in the cases of 
nonadiabatic H-resonance independent of mass hierarchy at 
$\varepsilon_\nu = 50$ MeV.
The same conclusion is inferred for neutrinos at different neutrino energies.

\subsection{Yields of $^7$Li and $^{11}$B}

In this subsection we discuss detailed dependence of the $^7$Li and $^{11}$B 
yields on the mass hierarchy and $\sin^{2}2\theta_{13}$.
When we do not consider neutrino oscillations, the calculated yields of
$^7$Li and $^{11}$B are $2.36 \times 10^{-7}$ and
$6.26 \times 10^{-7} M_\odot$.
We evaluate the ratios of the $^7$Li and $^{11}$B yields with and without 
neutrino oscillations.

Figure 4 shows the dependence of the yield ratios of $^7$Li ({\it a}) and 
$^{11}$B ({\it b}) on $\sin^{2}2\theta_{13}$ in the range  between
$1 \times 10^{-6}$ and $1 \times 10^{-1}$.
We find three characteristics of the $^7$Li and $^{11}$B
yields as to the dependence on $\sin^{2}2\theta_{13}$.
In the case of $\sin^{2}2\theta_{13} \la 2 \times 10^{-5}$, the $^7$Li
and $^{11}$B yield ratios keep constant values and do not depend on mass
hierarchy.
In the range of 
$2 \times 10^{-5} \la \sin^{2}2\theta_{13} \la 2 \times 10^{-3}$,
the $^7$Li and $^{11}$B yield ratios increase with $\sin^{2}2\theta_{13}$.
The difference due to mass hierarchy is also seen.
In the range of $\sin^{2}2\theta_{13} \ga 2 \times 10^{-3}$, the yield ratios
roughly keep constant values again.
The difference due to mass hierarchy is also seen in this range.
The above dependence is compared with that of the adiabaticity parameter 
$\gamma$ (see Eq. (10)) and the flip probability $P_f$ (see eq. (11))
of H-resonance at the location of the resonance density.
The value $1-P_f$ related to $\sin^{2}2\theta_{13}$ is also drawn in
Fig. 4.
We set the neutrino energy to be 50 MeV.
This neutrino energy is close to the optimum energy contributing most 
strongly to the charged-current $\nu$-process reactions of $^4$He.

In the case of $\sin^{2}2\theta_{13} \la 2 \times 10^{-5}$, $1-P_f$ is
almost equal to 0; H-resonance is nonadiabatic.
In the case of 
$2 \times 10^{-5} \la \sin^{2}2\theta_{13} \la 2 \times 10^{-3}$, $1-P_f$
increases with $\sin^{2}2\theta_{13}$, i.e., the resonance changes from
nonadiabatic to adiabatic with increasing $\sin^{2}2\theta_{13}$.
We call this range of $\sin^{2}2\theta_{13}$ ^^ ^^ transition range''.
In the case of $\sin^{2}2\theta_{13} \ga 2 \times 10^{-3}$, the value of
$1-P_f$ is almost equal to 1, i.e., H-resonance is adiabatic.
The change of the flip probability $P_f$ as a function of 
$\sin^{2}2\theta_{13}$ is roughly similar to the change of the yields of 
$^7$Li and $^{11}$B.
Thus, we conclude that the dependence of the $^7$Li and $^{11}$B yields
on $\sin^{2}2\theta_{13}$ strongly correlates to the adiabaticity of
H-resonance.

In the case of the nonadiabatic range,
$\sin^{2}2\theta_{13} \la 2 \times 10^{-5}$, the $^7$Li and $^{11}$B
yield ratios are about 1.1 and 1.02, respectively, and they are independent
of mass hierarchy.
In the limit of $\sin^{2}2\theta_{13}=0$, the yield ratios still stay near 
the above values.
We showed in \S3 that the transition probabilities do not depend on mass
hierarchy in nonadiabatic region.
This characteristics can also be seen in the $^7$Li and $^{11}$B yield ratios.
The yield ratios 1.1 and 1.02 are slightly larger than unity, which reflect
small enhancement of the mass fractions of $^7$Li and $^{11}$B in outer 
region of the He/C layer as shown in \S 4.1.

In the transition range, the yield ratios of $^7$Li and $^{11}$B increase 
with $\sin^{2}2\theta_{13}$ and a difference relating to mass hierarchy 
appears.
The increase in the $^7$Li yield is due to the enhancement of the $^7$Be
production through $^4$He($\nu_e,e^-p)^3$He($\alpha,\gamma)^7$Be.
The increase in the $^{11}$B yield also arises from the enhancement of $^7$Li 
yield by way of $^7$Be.

In the adiabatic range, the $^7$Li yield ratio depends on mass hierarchy: 
the $^7$Li yield ratio is 1.9 in the case of
normal mass hierarchy, and 1.3 in inverted mass hierarchy.
As shown in \S 4.1, the increase in the reaction rate of 
$^4$H($\nu_e,e^-p)^3$He raises the $^7$Be production in
normal mass hierarchy.
The increase in the rate of $^4$He($\bar{\nu}_e,e^+n)^3$H also
raises the $^7$Li production in inverted mass hierarchy.
The increase in the reaction rate of $^4$He($\nu_e,e^-p)^3$He in normal mass 
hierarchy is larger than that of $^4$He($\bar{\nu}_e,e^+n)^3$H in inverted mass
hierarchy.
In the case of $T_{\nu_e}=6$ MeV, the reaction rate of
$^4$He($\nu_e,e^-p)^3$He is larger than that of
$^4$He($\bar{\nu}_e,e^+n)^3$H.
Further, the difference between $T_{\nu_e}$ and $T_{\nu_{\mu,\tau}}$ 
is larger than that between $T_{\bar{\nu}_e}$ and $T_{\bar{\nu}_{\mu,\tau}}$
(where note that $T_{\nu_{\mu,\tau}} = T_{\bar{\nu}_{\mu,\tau}}$).
Thus, the increase in the $^7$Li yield ratio in normal mass hierarchy is 
larger than that in inverted mass hierarchy.
The $^{11}$B yield ratio also depends on mass hierarchy, but the difference is 
smaller than that of the $^7$Li yield ratio.
As mentioned above, the increase in the $^{11}$B yield arises in this case
from enhanced $^7$Li production by way of $^7$Be.
However, most of $^7$Li produced through $^7$Be do not capture 
$\alpha$-particles to form $^{11}$B.

\subsection{Dependence on Initial $\nu_e$ and $\bar{\nu}_e$ Temperatures}

In the present study we adopted the temperatures of $\nu_e$,
$\bar{\nu}_e$, and $\nu_{\mu,\tau}$ ($\bar{\nu}_{\mu,\tau}$) equal to
3.2 MeV, 5.0 MeV, and 6.0 MeV, respectively.
On the other hand, some other studies \citep[e.g.][]{ww95,rh02}
adopted the temperature of $\nu_e$ and $\bar{\nu}_e$
as $T_{\nu_e} = T_{\bar{\nu}_e} = 4.0$ MeV.
Since the enhancement of the $^7$Li and $^{11}$B yields depends on the
neutrino temperatures at the neutrino sphere, we can find some effects
on different temperatures of $\nu_e$ and $\bar{\nu}_e$.
We consider four sets of the temperatures of $\nu_e$ and 
$\bar{\nu}_e$:
($T_{\nu_e},T_{\bar{\nu}_e}$) = (3.2 MeV, 5 MeV), (4 MeV, 4 MeV),
(4 MeV, 5 MeV), (3.2 MeV, 4 MeV) as mentioned in \S 2.1.
The first set is our standard model.
The second set is the one adopted in \citet{ww95}, \citet{rh02}, and
so on.
The third set satisfy $T_{\nu_e} < T_{\bar{\nu}_e}$, but
the difference of the temperature is smaller than the first set.
The fourth  set is prepared for comparison to the other three models:
either $T_{\nu_e}$ or $T_{\bar{\nu}_e}$ is different from
the other sets.
In these four sets we use a temperature of $\nu_{\mu,\tau}$ and 
$\bar{\nu}_{\mu,\tau}$, 6.0 MeV.
The $^7$Li and $^{11}$B yields without neutrino oscillations
in the four cases of the neutrino temperatures are listed in Table 2.

Figure 5 shows the $^7$Li and $^{11}$B yield ratios as functions of 
$\sin^{2}2\theta_{13}$ in the above four cases of neutrino temperatures.
In the case of normal mass hierarchy (see Fig. 5{\it a}), the $^7$Li yield
ratio depends on initial temperatures in the nonadiabatic range, and the
dependence becomes larger in the adiabatic range.
In the case of small temperatures of $\nu_e$ and $\bar{\nu}_e$,
i.e., ($T_{\nu_e}, T_{\bar{\nu}_e}$)=(3.2 MeV, 4 MeV), the $^7$Li
yield ratio becomes 2.1 at the maximum.
On the other hand, large neutrino temperature, i.e.,
($T_{\nu_e}, T_{\bar{\nu}_e}$)=(4 MeV, 5 MeV), provides the
$^7$Li yield ratio of 1.8.
In the adiabatic range, the final $^7$Li yield does not depend on
the initial neutrino temperatures.
The difference is due to the difference of the $^7$Li yields without 
neutrino oscillations.
After the neutrino mixing at H-resonance, the $\nu_e$ energy spectrum
becomes FD distribution with $T_{\nu_e} = 6$ MeV, independent of the
initial $T_{\nu_e}$.
In addition, the yield produced originally as $^7$Be is much larger than the
$^7$Li yield produced without neutrino oscillations.
Therefore, the $^7$Be yield does not depend strongly on the neutrino 
temperatures among the four sets.

In the case of inverted mass hierarchy (see Fig. 5{\it b}), the dependence of 
the $^7$Li yield ratio on the initial neutrino temperatures
is similar to that in normal mass hierarchy.
The yield ratio is about 1.4 in the case of
($T_{\nu_e}, T_{\bar{\nu}_e}$)=(3.2 MeV, 4 MeV) and in the
adiabatic range.
It is 1.26 in the case of 
($T_{\nu_e}, T_{\bar{\nu}_e}$)=(4 MeV, 5 MeV).
In the adiabatic range the produced amount of $^7$Li through
$^4$He($\bar{\nu}_e,{\rm e^-}n)^3$H does not depend on the
initial neutrino temperatures.
The different neutrino temperatures lead to different $^7$Li yields through
$^4$He($\nu,\nu'p)^3$H, which makes virtually larger effect than the
neutrino oscillations.

Figure 5{\it c} shows the cases of the $^{11}$B yield ratios in the case of 
normal mass hierarchy.
The $^{11}$B yield ratios are 1.31 and 1.23 in the cases of 
($T_{\nu_e},T_{\bar{\nu}_e}$)=(3.2 MeV, 4 MeV) and 
(4 MeV, 5 MeV), respectively, in adiabatic range.
Since most of $^{11}$B is produced by way of $^7$Li, the dependence on
the neutrino temperatures is similar to $^7$Li.
Some $^{11}$B are produced by way of $^7$Be through 
$^7$Be($n,p)^7$Li($\alpha,\gamma)^{11}$B.
Thus, the difference of $^{11}$B yield due to different neutrino temperatures
is smaller in the adiabatic case than in the nonadiabatic case.
From the viewpoint of the yield ratios, the $^{11}$B yield ratio is larger
in the adiabatic range.

In the case of inverted mass hierarchy (see Fig. 5{\it d}), the $^{11}$B 
yield ratios are 1.39 and 1.21 in the cases of 
($T_{\nu_e}, T_{\bar{\nu}_e}$)=(3.2 MeV, 4 MeV) and
(4 MeV, 5 MeV) in the adiabatic range.
In the case of ($T_{\nu_e},T_{\bar{\nu}_e}$)=(3.2 MeV, 5 MeV)
the $^{11}$B yield ratio is 1.35 in the adiabatic range.
Thus, the $^{11}$B yield ratios in the cases of $T_{\nu_e} = 3.2$ MeV
are larger than the maximum $^{11}$B yield ratio in normal mass hierarchy.
In these cases the $^7$Li yield ratios are correspondingly similar to each
other.
This reflects the fact that $^{11}$B is mainly produced by way of $^3$H and 
$^7$Li.

\section{Discussion}

\subsection{Neutrino Oscillations in Shock Propagating Medium}

When we calculated neutrino oscillations, we adopted the density profile
of a presupernova and did not consider the shock propagation during supernova
explosion.
One of the reasons is that our hydrodynamical model assumes the inner 
boundary of the supernova ejecta as a mass cut and the density 
structure inside the mass cut is not considered.
Indeed, numerical simulations of core collapse, core bounce, and explosion
of surrounding materials have not been successful yet \citep{br03,ss04}.
On the other hand, numerical simulations of the shock propagation from
a neutrino sphere to the envelope have been studied using time-dependent
inner boundary, i.e., the surface of the neutron star \citep[e.g.,][]{tk04}.
In this section we discuss the shock propagation effect on neutrino 
oscillations using the density profile of the supernova ejecta calculated 
by hydrodynamical model, and assuming a simple analytical density profile 
inside the mass cut.

In order to discuss the shock propagation effect on the neutrino oscillations,
we take simple analytical density structure inside the mass cut.
First, we assume that the density drops as $\rho \propto r^{-3}$ from 
a neutrino sphere until it drops to the value at the mass cut.
Then, the density stays a constant at the value for the location
of the mass cut.
The density profile in the present discussion is thus
\begin{equation}
\rho(r) = 1 \times 10^{4} \left( \frac{r}{1 \times 10^9 {\rm cm}} \right)^{-3}
\quad {\rm for} \quad r \le r_{\rm c}
\end{equation}
and
\begin{equation}
\rho(r) = \rho_{{\rm mcut}} 
\quad {\rm for} \quad r_{\rm c} \le r \le r_{\rm mcut},
\end{equation}
where $\rho_{\rm mcut}$ and $r_{\rm mcut}$ are the density and radius
at the mass cut and $r_{\rm c}$ is the radius where the density first drops to
$\rho = \rho_{{\rm mcut}}$.
As the shock wave moves outward, the density behind the shock front
decreases and the density at the mass cut also drops.
Thus, $r_{\rm c}$ becomes larger as time passes by.

We now study the influence of the neutrino oscillations with the shock
propagation to nucleosynthesis of the light elements.
Figure 6 shows the dependence of the $^7$Li and $^{11}$B yield ratios on
$\sin^{2}2\theta_{13}$.
First, we consider the dependence in the normal mass hierarchy.
We find in the case of $\sin^{2}2\theta_{13} \ga 1 \times 10^{-4}$ that 
the yield ratios of both $^7$Li and $^{11}$B are slightly smaller than
the corresponding yield ratios in the presupernova density profile without
shock propagation effect.
The decrease in the yield ratios is about at most 4\% in the case of
$\sin^{2}2\theta_{13} \sim 1 \times 10^{-3}$.
The shock propagation effect of the neutrino oscillations is seen in the 
case of H-resonance close to adiabatic.

Next, we consider the case of inverted mass hierarchy.
Compared with the case of normal mass hierarchy, the decrease in the yield
ratios are smaller; they are 1\% level in the case of 
$\sin^{2}2\theta_{13} \sim 1 \times 10^{-3}$.
Therefore, we can conclude that in the inverted mass hierarchy not only 
the effect of neutrino oscillations itself but also the shock propagation 
effect is smaller than the case of normal mass hierarchy.

The shock propagation effect on the neutrino oscillations is seen in
$5 \sim 10$ s when the shock front reaches the O/C layer or the inner region 
of the He/C layer.
Meantime, the shock wave passes through the H-resonance region.
When the shock wave passes through this region, the density becomes higher
and the resonance region moves outward in the mass coordinate.
At the same time, the density change at the shock front makes the 
adiabaticity of the resonance slightly smaller.
After the shock front has passed the O/C layer, the location of the 
resonance in the mass coordinate goes inward owing to the density decrease
by the expansion.
The shock propagation effect is mainly seen in the inner region of the
He layer, $3.8 M_\odot \la M_r \la 4.3 M_\odot$.
We have shown in the last section that $^7$Be is produced in the region
of $M_r \la 4.8 M_\odot$ and that $^7$Li is produced outside this region.
In the normal mass hierarchy the increase in the yield is mainly
due to $^7$Be production.
Therefore, the shock propagation effect is seen more clearly in the normal 
mass hierarchy.

\subsection{Dependence on Stellar Models}

We investigated the $\nu$-process nucleosynthesis with neutrino oscillations
using 14E1 model, which is a progenitor model of SN 1987A in \citet{sn90}.
This stellar model corresponds to about 20 $M_\odot$ at the zero-age main 
sequence (ZAMS) and has a He core of 6.0 $M_\odot$.
Stars with the mass larger than $\sim 12 M_\odot$ are considered to evolve
to form Fe core and become core-collapse supernovae at the end.
Their internal structure indicates ^^ ^^ onion'' shell structure and their 
abundance distribution depends on their stellar mass at the ZAMS.
On the other hand, the treatment of convection also affects the structure
of presupernovae.
We here discuss the influence of the internal structure to the $\nu$-process
nucleosynthesis with neutrino oscillations.

Detailed studies of massive star evolution \citep[e.g.,][]{nh88,ha95} 
indicated that the region of the O-rich layers in the presupernova stage 
increases in larger ZAMS stellar masses.
\citet{ha95} indicated that 4, 8, and 16 $M_\odot$ He star models have
the O-rich layers of 0.7, 4.4, and 11.4 $M_\odot$ and that the region
of the He-rich layers has commonly about masses of $\sim 2 M_\odot$.
Stellar models including semiconvection and convective overshooting mixing 
also  have large O-rich layers \citep{ww88}.
If the region of the O-rich layers is much larger than that of the He/C 
layer, most of $^{11}$B would be produced through the $\nu$-process from
$^{12}$C in the O-rich layers rather than from $^4$He in the He/C layer.
On the other hand, the density in the O/C layer scarcely depends on the
stellar mass \citep[e.g.,][]{nh88,ha95} and the density is close to the
density of H-resonance.
We showed that the effect of neutrino oscillations is not seen in the 
O-rich layers.
Thus, we expect that the increase in the $^{11}$B yield due to neutrino
oscillations is very small if the region of the O-rich layers is larger 
than the He/C layer and $^{11}$B is mainly produced from $^{12}$C.

The density profile of the He/C layer also may affect the increase in
the $^7$Li yield by neutrino oscillations.
This is because $^7$Li is produced as $^7$Be in the inner region
of the He/C layer and as $^7$Li outside the region.
We expect that stellar models which produce more $^7$Be rather than 
$^7$Li bring about larger increase in the $^7$Li yield due to neutrino 
oscillations in adiabatic H-resonance and normal mass hierarchy.
In our model, the yield of $^7$Be increases by a factor of 3.1 at the maximum,
whereas $^7$Li produced through $^3$H($\alpha,\gamma)^7$Li increases
by a factor of 1.2 in the normal mass hierarchy.
When we consider adiabatic H-resonance and normal mass hierarchy, 
stellar models that produce $^7$Li as $^7$Be rather than $^7$Li
lead to the $^7$Li yield more than twice as that without oscillations.
On the other hand, if the contribution from $^7$Li through 
$^3$H($\alpha,\gamma)^7$Li is large, the increase in the $^7$Li yield due to 
the oscillation would be smaller in the normal mass hierarchy,
and it would be larger in the inverted mass hierarchy.
In the latter case, the $^7$Li yield produced through 
$^3$H($\alpha,\gamma)^7$Li increases by a factor of 1.5 at the maximum.

The shock propagation effect on the neutrino oscillations would depend
on the stellar mass.
Less massive stars have smaller O-rich region, so that it takes shorter
time until the shock wave arrive at the H-resonance density region.
After the shock arrival, the transition probabilities to the other flavors
become small and the increase in the rates of charged-current $\nu$-process
reactions also reduces.
Thus, the increase in the $^7$Li and $^{11}$B yields is expected to be small
for supernovae evolved from less massive stars.

\subsection{$CP$ Phase Effect}

In this study, we assume that the $CP$ phase $\delta$ is equal to zero 
because the definite value of $\delta$ has not been determined
from neutrino experiments.
Let us briefly discuss the influence of $CP$ phase on the 
$^7$Li and $^{11}$B yields produced through the $\nu$-process.
We consider the transition probabilities to $\nu_e$ or
$\bar{\nu}_e$.
\citet{yk02} showed exact relations of neutrino transition probabilities in
arbitrary matter profile.
They showed that the transition probability of 
$\nu_e \rightarrow \nu_e$
($\bar{\nu}_e \rightarrow \bar{\nu}_e$) does not depend on the
$CP$ phase $\delta$.
From their study, we also obtained that the sum of the transition
probabilities of $\nu_\mu \rightarrow \nu_e$ and
$\nu_\tau \rightarrow \nu_e$
($\bar{\nu}_\mu \rightarrow \bar{\nu}_e$ and
$\bar{\nu}_\tau \rightarrow \bar{\nu}_e$) does not depend on the
$CP$ phase.
Here, we note that $\nu_\mu$ and $\nu_\tau$ 
($\bar{\nu}_\mu$ and $\bar{\nu}_\tau$) emitted from a proto-neutron
star have a same energy spectrum, i.e., the numbers of $\nu_\mu$ and 
$\nu_\tau$ ($\bar{\nu}_\mu$ and $\bar{\nu}_\tau$) are same for a given
neutrino energy.
Since the numbers of $\nu_\mu$ and $\nu_\tau$ 
($\bar{\nu}_\mu$ and $\bar{\nu}_\tau$) for a given neutrino energy are
same, the transition probability from $\nu_\mu$ or $\nu_\tau$ 
($\bar{\nu}_\mu$ or $\bar{\nu}_\tau$) to $\nu_e$ ($\bar{\nu}_e$) for a given 
neutrino energy does not depend on the $CP$ phase, and therefore, the change 
of the energy spectrum of $\nu_e$ ($\bar{\nu}_e$) does not depend on the
$CP$ phase, too.
We do not expect any influences of the $CP$ phase to the $^7$Li and $^{11}$B 
yields.

\subsection{Active-sterile neutrinos}

It is noted that the existence of the fourth sterile neutrinos may change
the effect of supernova light element synthesis on neutrino oscillations.
\citet{fm03} considered additional sterile neutrinos with the mass squared
difference of $1 \sim 100$ eV$^2$ from electron neutrinos and investigated 
the effect of active-sterile neutrino conversion on $r$-process in 
neutrino-driven wind model.
The Los Alamos liquid scintillator neutrino detector (LSND) experiment 
suggested that neutrino oscillations occur with mass squared difference 
in the range of $\Delta m^2 = 0.2 \sim 10$ eV$^2$ \citep{LSND01}.
They argued that it is difficult to generate neutrons enough to activate
the $r$-process.
Then, they showed that active-sterile neutrino conversion makes the winds
favorable for the $r$-process even in the initial condition unfavorable to
the $r$-process.
This is because the active-sterile neutrino conversion ceases $\nu_e$
and, thus, the neutron depletion by $n(\nu_e,e^-)p$ is suppressed.

If the active-sterile neutrino conversion occurs effectively just after
the onset of supernova explosion, then this conversion may reduce the number 
of $\nu_e$ and, therefore, it may reduce $^7$Li and $^{11}$B yields.
It is expected that the active-sterile neutrino conversion occurs in inner 
region of the O-rich layer or even deeper region 
($\rho \sim 6 \times 10^5$ g cm$^{-3}$ for $\Delta m^2 = 1$ eV$^2$ and 
$\varepsilon_\nu = 20$ MeV; see Eq. (10)).
The decrease in the number of $\nu_e$ may reduce not only the yields
of $^7$Li and $^{11}$B but also those of heavy neutron-deficient nuclei
like $^{138}$La and $^{180}$Ta more effectively, which are produced through
$(\nu_e,e^-)$ reactions.
In practice, the effect of the active-sterile neutrino conversion is still 
complicated because mixing angles of active-sterile neutrinos have not been 
determined and the conversion may strongly depend on the mixing angles.

\subsection{Prospects for Constraining Mass Hierarchy and $\theta_{13}$}

We obtained that the $^7$Li yield is larger by about a factor of two by
taking account of neutrino oscillations of normal mass hierarchy and 
adiabatic H-resonance compared with that without neutrino oscillations.
If we can detect the $^7$Li abundance or the abundance ratio of $^7$Li to 
an element that is not affected by the $\nu$-process in stars which show clear
traces of supernovae and if one compares them with the evaluated abundance 
or abundance ratio in supernova ejecta as we discussed, we may be able to 
constrain mass hierarchy and the mixing angle $\theta_{13}$.
We show the dependence of the number ratio of $^7$Li/$^{11}$B on 
$\sin^{2}2\theta_{13}$ in Fig. 7.
The $^7$Li/$^{11}$B ratio should be better than the abundance of $^7$Li itself
for observations because the $^7$Li/$^{11}$B ratio is rather insensitive to 
the uncertainties of supernova neutrinos \citep{yk06}.
When we evaluated the number ratio of $^7$Li/$^{11}$B, we considered the four 
sets of the temperatures of $\nu_e$ and $\bar{\nu}_e$ discussed in \S 4.3.
For all three ranges of normal and inverted mass hierarchies and without
neutrino oscillations, the largest and smallest values correspond to the 
cases of ($T_{\nu_e}, T_{\bar{\nu}_e}$) = (3.2 MeV, 4.0 MeV)
and (4.0 MeV, 5.0 MeV).
The $^7$Li/$^{11}$B ratio with ($T_{\nu_e}, T_{\bar{\nu}_e}$) = 
(4.0 MeV, 4.0 MeV) is larger than the result with the neutrino temperatures
in our standard model.
In the normal mass hierarchy and 
$\sin^{2}2\theta_{13} \ga 2 \times 10^{-3}$, the $^7$Li/$^{11}$B ratio is 
larger than 0.83.
The maximum values of the $^7$Li/$^{11}$B ratio in the inverted 
mass hierarchy and without neutrino oscillations are 0.71 and 0.63.
We note that the range of the $^7$Li/$^{11}$B ratio deduced from the 
uncertainties of the $\nu_{\mu,\tau}$ temperatures, which have been 
investigated in \citet{yk06}, is included in the range obtained in the
present study.
Therefore, the $^7$Li/$^{11}$B number ratio in the normal mass hierarchy and 
adiabatic H-resonance is larger than that without the oscillations.
The $^7$Li/$^{11}$B ratio could be a tracer of normal mass hierarchy and 
relatively larger $\theta_{13}$, still satisfying the observational
constraint of $\sin^{2}2\theta_{13} \le 0.1$, if the number ratio of 
$^7$Li/$^{11}$B is precisely observed.

Recently, measurements of B isotopic ratio in metal poor stars have been
challenged by several groups \citep[e.g.,][]{rd98,rp00}.
Since supernovae provide large amount of $^{11}$B with small amount of
$^{10}$B, the stars having traces of a supernova are expected to indicate 
$^{11}$B/$^{10}$B ratio larger than that of the solar-system composition.
Deficiency of B abundance with normal primordial $^7$Li is reported in
a metal poor star formed in an epoch when supernova nucleosynthesis dominated
in the early Galaxy with quenched contribution from the cosmic ray 
interactions with interstellar medium \citep{pd98,pd99}.
The ratio of $^7$Li/$^{11}$B in supernovae might be found also in primitive
meteorites.

Primitive meteorites should contain supernova-originating materials
and $^7$Li/$^{11}$B ratio might constrain the neutrino 
oscillation parameters.
The solar-system $^7$Li/$^{11}$B ratio is 3.11 \citep{ag89}.
The $^7$Li/$^{11}$B in the Galactic cosmic rays, which is one of the main
production sites of Li and B, is evaluated to be 0.58 \citep{fo00}.
These two values are out of our $^7$Li/$^{11}$B range evaluated in the 
supernova $\nu$-process.
Therefore, supernova-originating material should have $^7$Li/$^{11}$B ratios
different from those of the solar-system composition and the Galactic cosmic
rays, which is a favorable feature for our detectable $^7$Li/$^{11}$B
ratio in the range $0.6 \la N(^7{\rm Li})/N(^{11}{\rm B}) \la 1.0$ .
Recently, presolar grains from supernovae have been found and isotopic ratios
of C, N, O, Si, and Ti have been measured \citep[e.g.][]{ah92,na96}.
Measurement of B isotopic ratios in the grains has been attempted \citep{hl01}.
Future isotopic and abundance measurements of Li and B would deduce 
$^7$Li/$^{11}$B ratio in supernovae.

We have to note, however, that there are many uncertainties to evaluate
the $^7$Li/$^{11}$B ratio and the Li yield in supernova ejecta.
These values depend on the stellar mass at least theoretically.
As discussed in \S5.2, treatment of convection may also change stellar 
structure and abundance distribution.
Aspherical explosion may produce different amount of light elements.
In our study, the energy dependence of $\nu$-process reaction cross sections
is very simplified: more precise evaluation of the energy dependence is 
required \citep{sc06}.
In order to evaluate the enhancement factor due to neutrino oscillations,
we need to construct more precise nucleosynthesis model in massive star
evolution and supernova explosions.
We also need to observe $^7$Li/$^{11}$B ratio and Li abundance in stars 
having traces of a definite supernova.

We showed the possibility of constraining parameters of neutrino oscillations 
from the viewpoint of nucleosynthesis.
The enhancement of $^7$Li/$^{11}$B ratio and $^7$Li abundance would be seen 
in the normal mass hierarchy and adiabatic H-resonance.
There are different approaches to constrain neutrino oscillation parameters.
One is to constrain neutrino masses from cosmological observations.
Massive neutrinos produced in the early universe affect Cosmic Microwave
Background (CMB) power spectrum and structure formation, i.e., the shape of
matter power spectrum \citep[review in][]{el05}.
Constraining the total mass of mass eigenstates of neutrinos
($m_{\nu,{\rm tot}} = m_1+m_2+m_3$) from cosmological observations has been
carried out.
An upper limit of the total neutrino mass, $m_{\nu,{\rm tot}} < 0.42$ eV 
at 95\% C.L., has been deduced by fitting cosmological
parameters combined with the analyses of Wilkinson Microwave Anisotropy
Probe (WMAP) CMB, galaxy clustering, galaxy bias, and Ly$\alpha$ forest
of the Sloan Digital Sky Survey \citep{sm05}.
In the near future higher precision cosmological observations would reduce 
upper limit of the total neutrino mass, from which the squared mass 
difference and mass hierarchy also would be determined.

Observations of Diffuse Supernova Neutrino Background (DSNB) may constrain
parameters of neutrino oscillations, too.
Although DSNB has not been detected yet, future improved neutrino detectors 
may detect DSNB.
\citet{ya05} proposed a gadolinium-enhanced Super-Kamiokande detector
by which average $\bar{\nu}_e$ energy and the total $\bar{\nu}_e$
energy per supernova are to be measured after five years run.
Further, \citet{an04} discussed that, in the inverted mass hierarchy
and adiabatic H-resonance, the average $\bar{\nu}_e$ energy is very
different from the expected one without the neutrino oscillations.
Therefore, combined with theoretical evaluation of the average 
$\bar{\nu}_e$ and $\bar{\nu}_{\mu,\tau}$ energies emitted from
proto-neutron stars, the effect of the neutrino oscillations will be
studied and the neutrino oscillation parameters will be constrained.

We have these three different procedures to constrain neutrino oscillation 
parameters.
First, the investigation of the $\nu$-process nucleosynthesis in supernovae
would provide a piece of evidence for normal mass hierarchy and adiabatic 
H-resonance.
Second, DSNB measurement with improved neutrino detector would reveal 
a possible proof of inverted mass hierarchy and adiabatic H-resonance.
Third, improved cosmological observations would clarify mass hierarchy.
Thus, combining the three investigations with each other, we hope to 
constrain strictly the neutrino oscillation parameters in the future.

\section{Conclusions}

We studied light element nucleosynthesis through the $\nu$-process
in supernovae taking account of neutrino oscillations.
The parameters of neutrino oscillations were adopted from the evaluations
through several neutrino experiments.
We used a supernova model corresponding to SN 1987A and investigated the
dependence of the $^7$Li and $^{11}$B yields on the mixing angle $\theta_{13}$
and mass hierarchy.
The obtained results are summarized as follows:
\begin{itemize}
\item
Neutrino oscillations affect the yields of $^7$Li and $^{11}$B synthesized
in supernova explosions.
Especially, the $^7$Li yield increases by a factor of 1.9 in the normal mass 
hierarchy and adiabatic H-resonance 
($\sin^{2}2\theta_{13} \ga 2 \times 10^{-3}$) compared with that without
neutrino oscillations.
The $^{11}$B yield increases by a factor of 1.3.

\item
In the inverted mass hierarchy, the increase in the $^7$Li and 
$^{11}$B yields is smaller: the yields of $^7$Li and $^{11}$B increase
by factors of 1.3 and 1.2.

\item
Neutrino oscillations in supernovae make the reaction rates of charged-current
$\nu$-process reactions larger.
The reaction rates of neutral-current $\nu$-process reactions do not change.
Thus, the final amounts of the $\nu$-process products increase by the 
neutrino oscillations.
In our study, main important $\nu$-process reactions for the $^7$Li and 
$^{11}$B production are $^4$He($\nu,\nu'p)^3$H($\alpha,\gamma)^7$Li, 
$^4$He($\nu,\nu'n)^3$He($\alpha,\gamma)^7$Be,
$^{12}$C($\nu,\nu'p)^{11}$B, and $^{12}$C($\nu,\nu'n)^{11}$C.
When we consider neutrino oscillations, the following charged-current
$\nu$-process reactions become also important:
$^4$He($\nu_e,e^-p)^3$He($\alpha,\gamma)^7$Li, 
$^4$He($\bar{\nu}_e,e^+n)^3$H ($\alpha,\gamma)^7$Be,
$^{12}$C($\nu_e,e^-p)^{11}$C,
$^{12}$C($\bar{\nu}_e,e^+n)^{11}$B.

\item
The neutrino temperatures also affect the $^7$Li and $^{11}$B
yields due to neutrino oscillations.
Large difference of the temperatures of $\nu_e$ and $\nu_{\mu,\tau}$
brings about larger increase in the yields compared with those without
neutrino oscillations.

\item
The shock propagation effect on the neutrino oscillations would slightly 
reduce the increment of the $^7$Li and $^{11}$B yields.
In our model, most of neutrinos pass through the He/C layer before the
shock wave arrives at the O/C layer, i.e., the resonance density region.
When the shock wave is in the O-rich layers,  the density change by the
shock does not influence strongly neutrino oscillations.

\end{itemize}

\acknowledgments

We would like to thank Koichi Iwamoto, Ken'ichi Nomoto, and Toshikazu 
Shigeyama for providing the data for the internal structure of progenitor
model 14E1 and for helpful discussions.
We are also indebted to Yong-Yeon Keum and Masahiro Takada for their 
valuable discussions.
Numerical computations were in part carried out on general common use computer
system at Astronomical Data Analysis Center (ADAC) of National
Astronomical Observatory of Japan. 
This work has been supported in part by the Ministry of Education, Culture, 
Sports, Science and Technology, Grants-in-Aid for Young Scientist (B) 
(17740130) and Scientific Research (17540275), for Specially Promoted
Research (13002001), and Mitsubishi Foundation.
T.Y. has been supported by the 21st Century COE Program 
^^ ^^ Exploring New Science by Bridging Particle-Matter Hierarchy'' in 
Graduate School of Science, Tohoku University.

\clearpage



\begin{figure}
\epsscale{.60}
\plotone{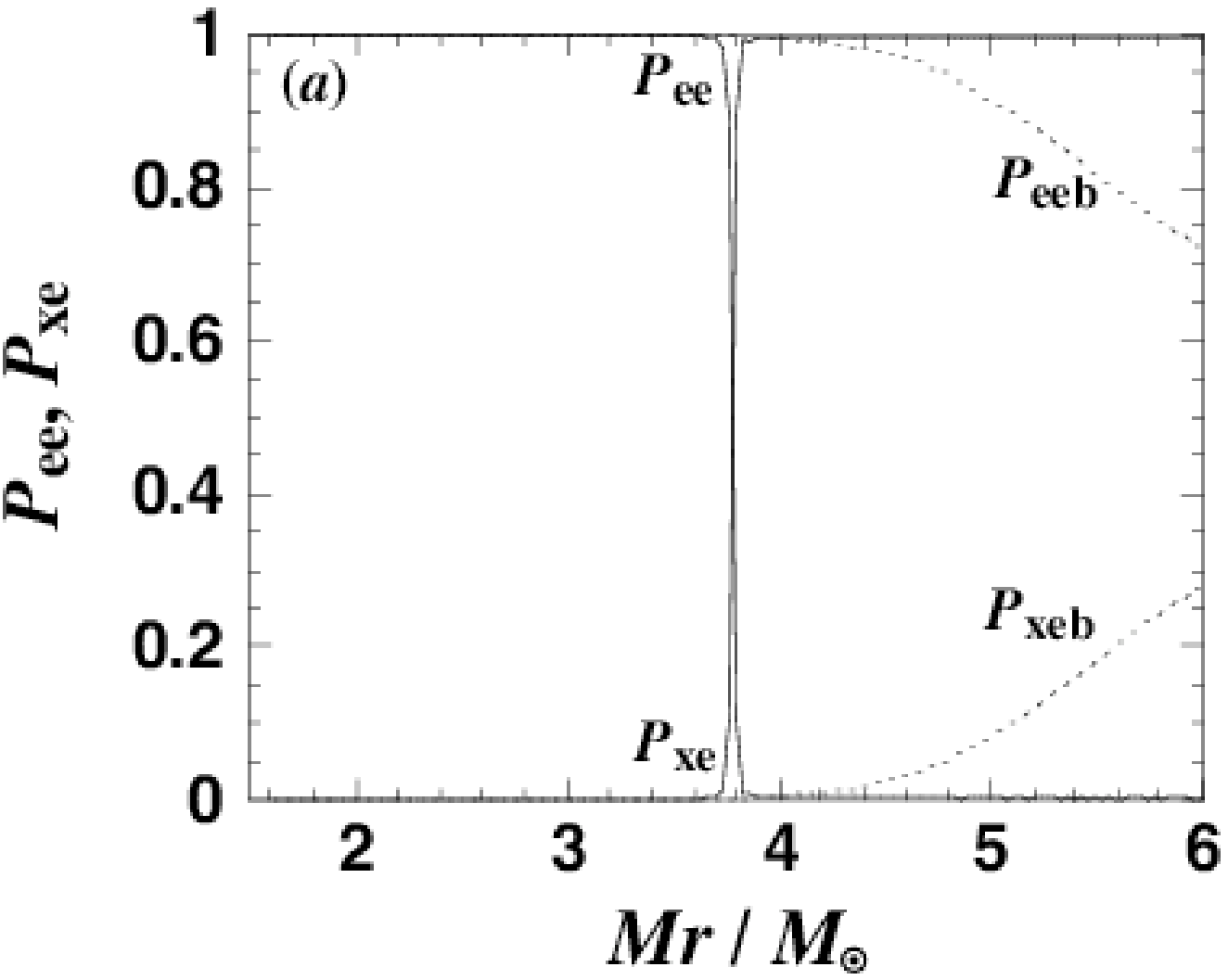}
\plotone{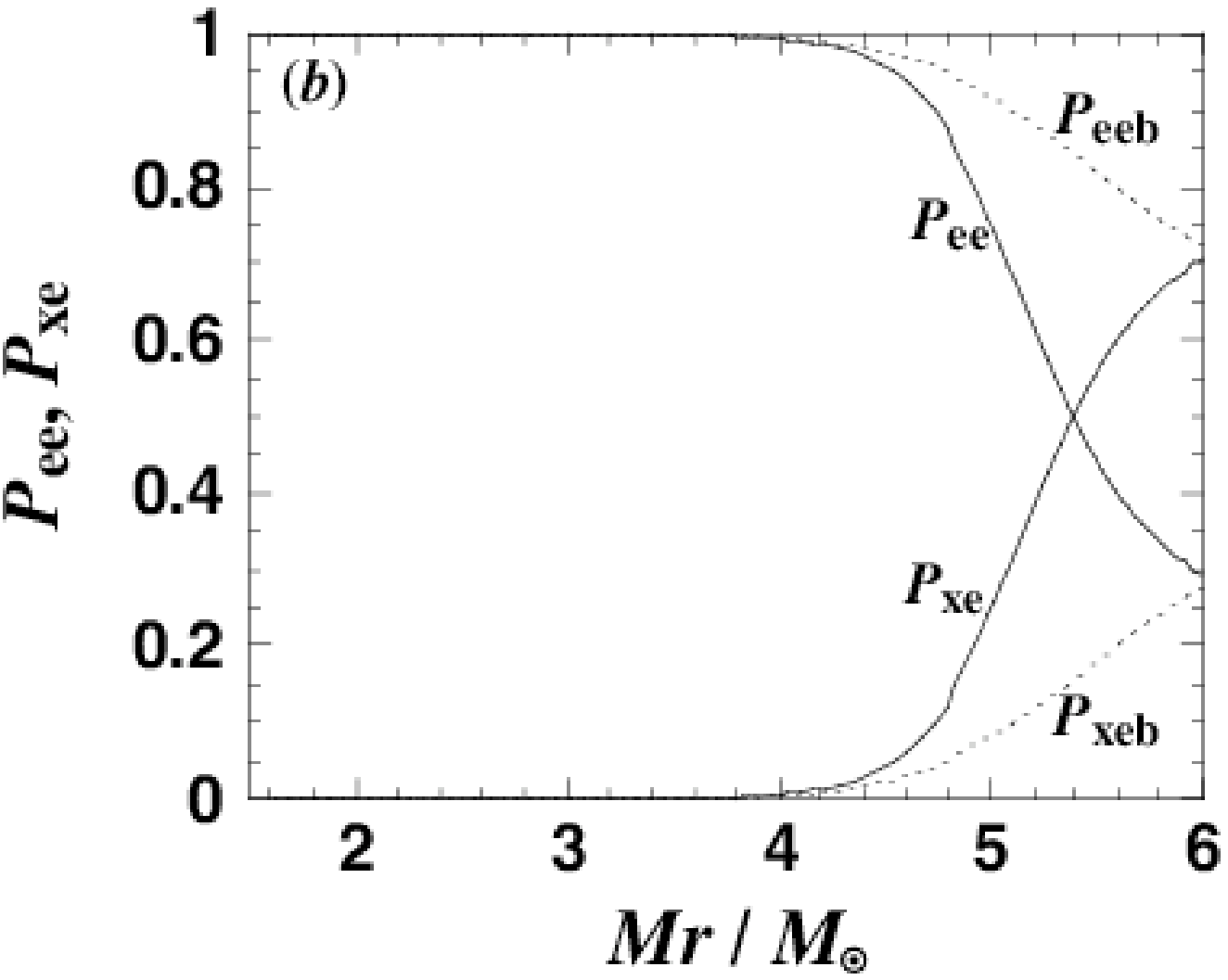}
\caption{
Transition probabilities by neutrino oscillations in the normal mass hierarchy.
The value of $\sin^{2}2\theta_{13}$ is ({\it a}) $1 \times 10^{-2}$ and 
({\it b}) $1 \times 10^{-6}$.
Solid lines are the transition probabilities from $\nu_e$ to 
$\nu_e$, $P_{\rm ee}$, and from $\nu_{\mu,\tau}$ to $\nu_e$
,$P_{\rm xe}$.
Dashed lines are those from $\bar{\nu}_e$ to $\bar{\nu}_e$,
$P_{\rm eeb}$, and from $\bar{\nu}_{\mu,\tau}$ to $\bar{\nu}_e$,
$P_{\rm xeb}$.
Horizontal axis is the mass coordinates in units of $M_\odot$.
The neutrino energy is set to be 50 MeV.
}
\end{figure}

\clearpage


\begin{figure}
\epsscale{0.6}
\plotone{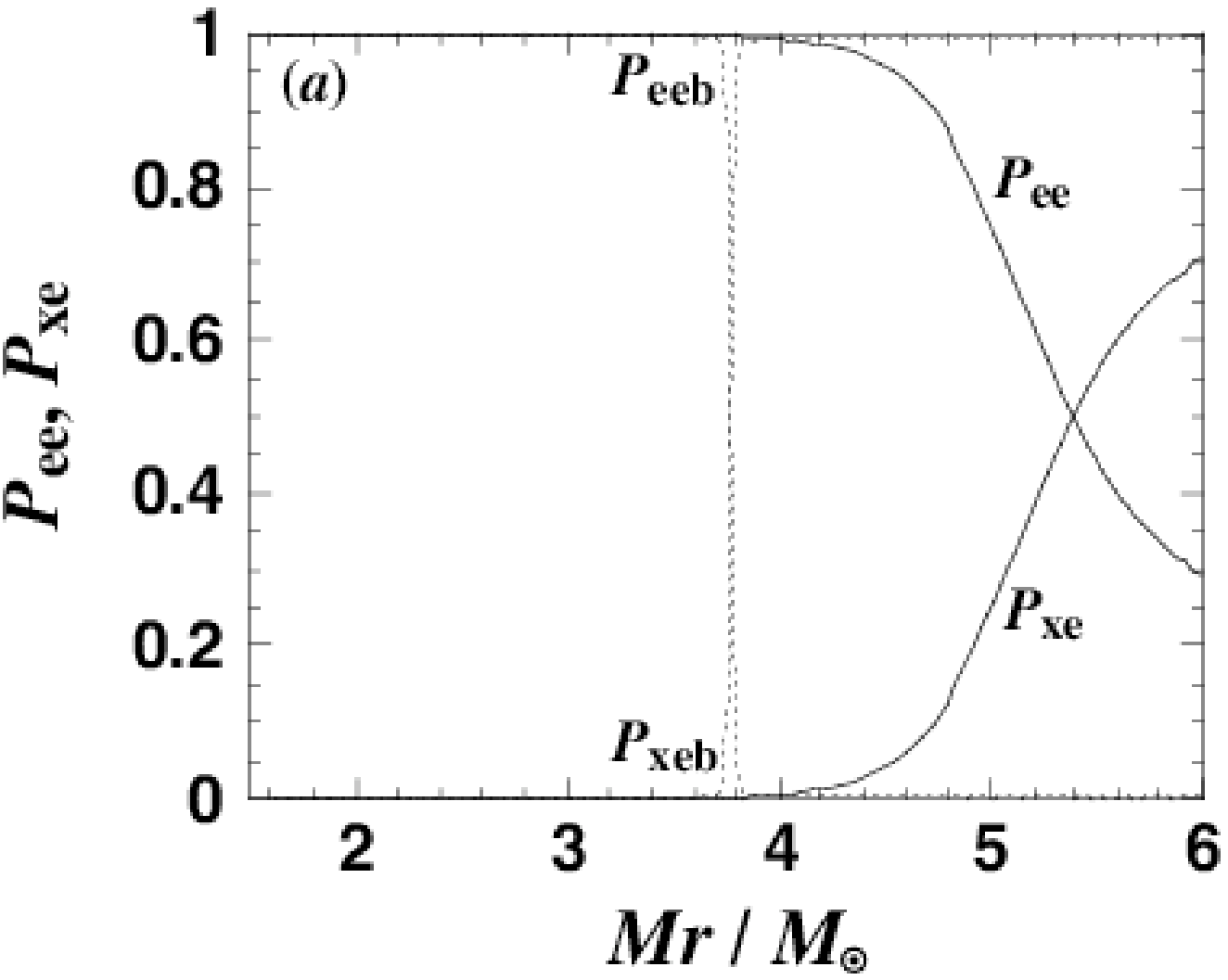}
\plotone{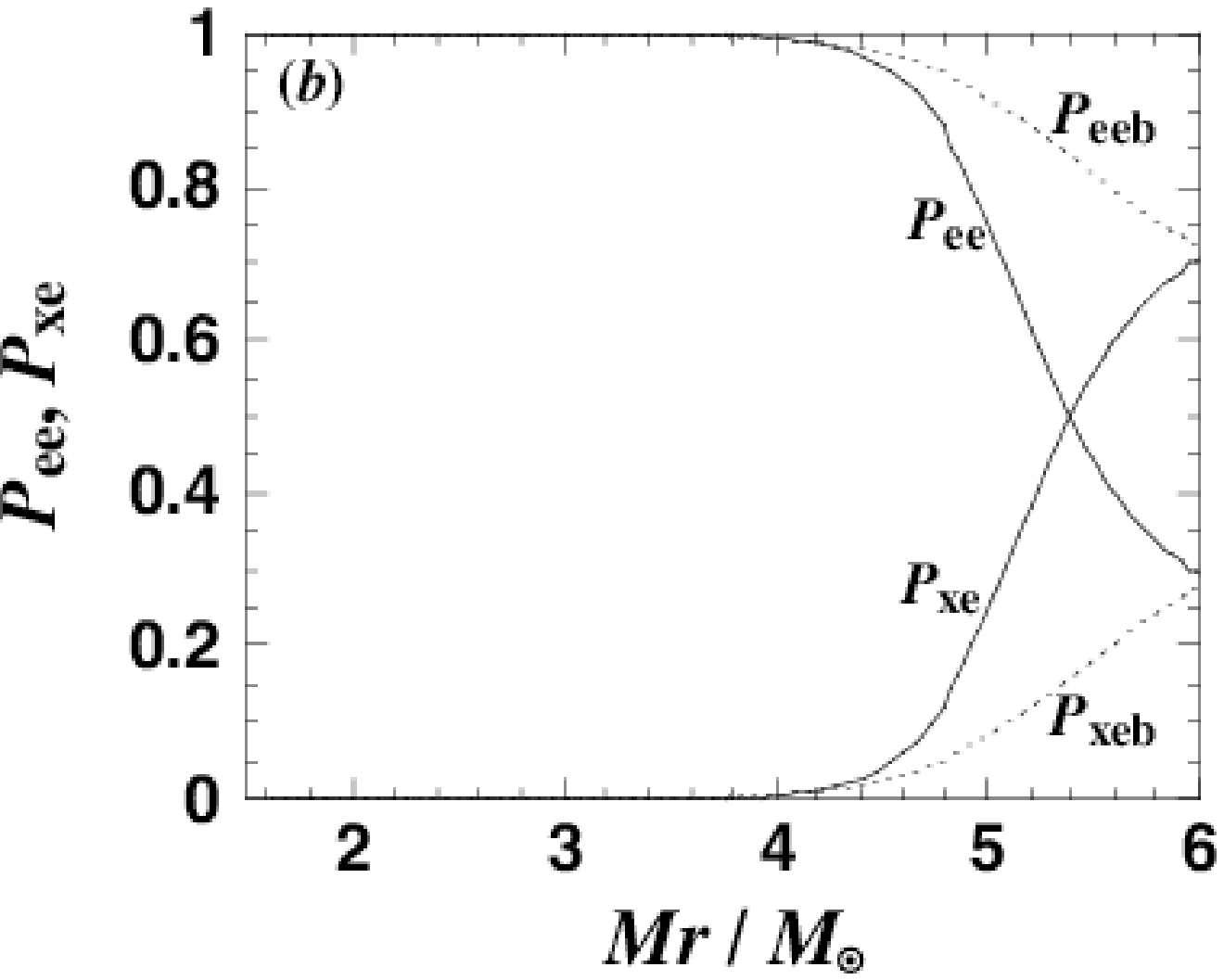}
\caption{
The same as Fig. 1 but for the inverted mass hierarchy.
}
\end{figure}

\clearpage
\begin{figure}
\epsscale{0.65}
\plotone{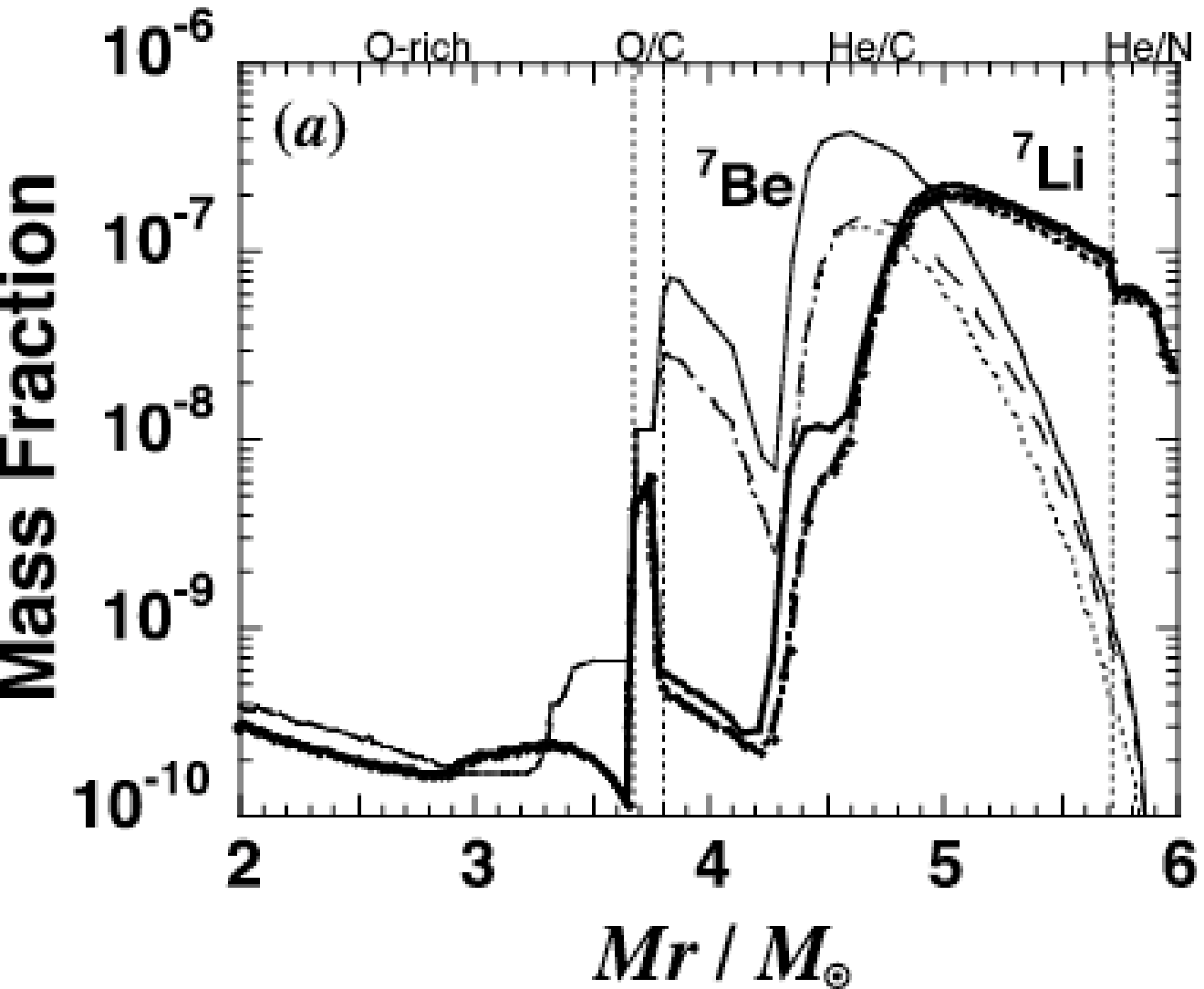}
\plotone{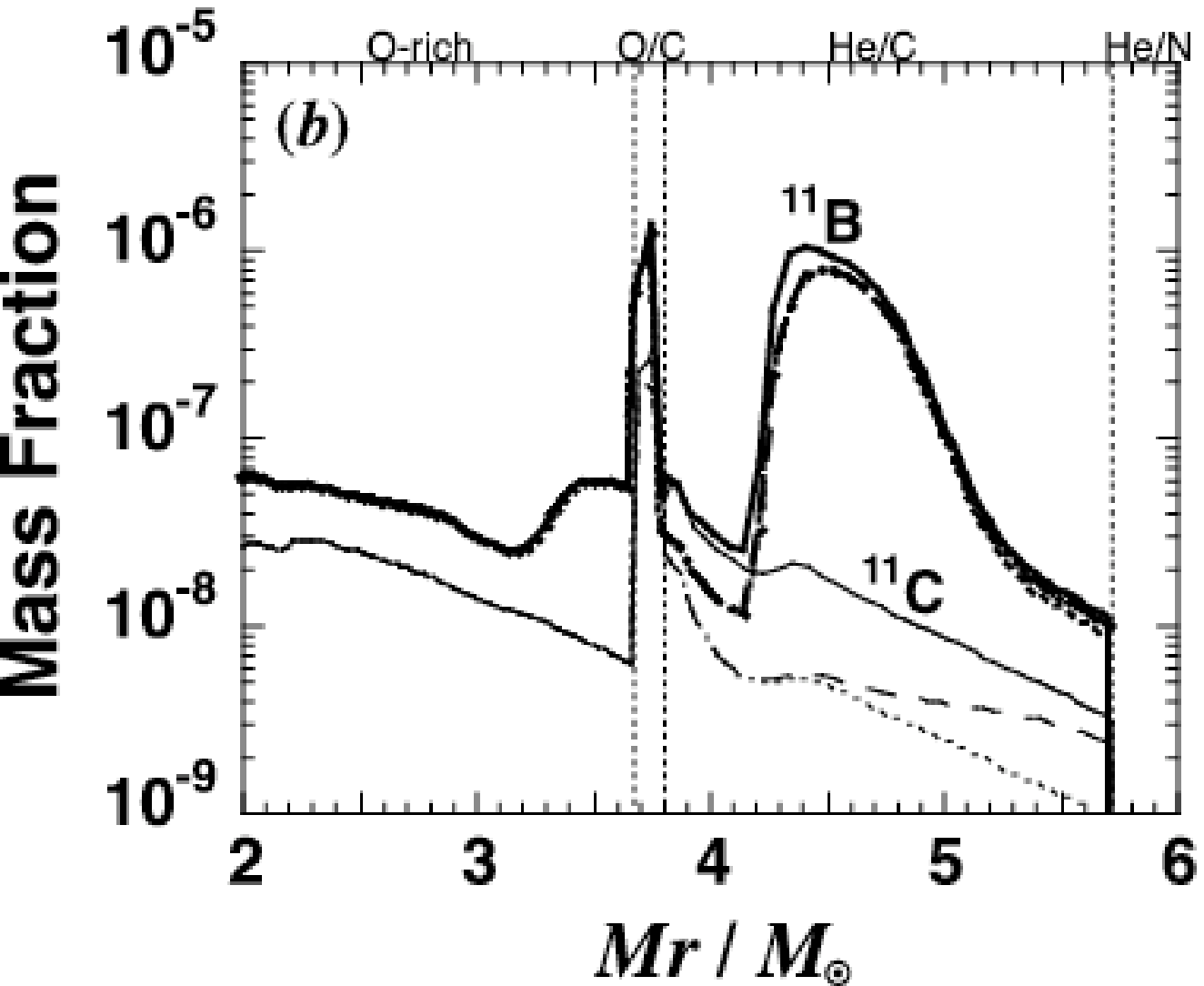}
\caption{
Mass fraction distribution of $^7$Li and $^{11}$B in the normal mass 
hierarchy (({\it a}) and ({\it b})) and inverted mass hierarchy 
(({\it c}) and ({\it d})) as a function of the mass coordinate in units
of $M_\odot$.
O-rich, O/C, He/C, and He/N indicate the O-rich, O/C, He/C, and He/N layers
in the supernova ejecta.
Thick lines and thin lines correspond to the mass fractions of $^7$Li and 
its isobar $^7$Be, respectively, in ({\it a}) and ({\it c}), and those of 
$^{11}$B and its isobar $^{11}$C, respectively, in ({\it b}) and ({\it d}).
Solid lines and dashed lines correspond to the cases of 
$\sin^{2}2\theta_{13} = 1 \times 10^{-2}$ and $1 \times 10^{-6}$.
The mass fractions calculated without neutrino oscillations are indicated
by dotted lines.
}
\end{figure}

\clearpage

\begin{figure}
\figurenum{3}
\plotone{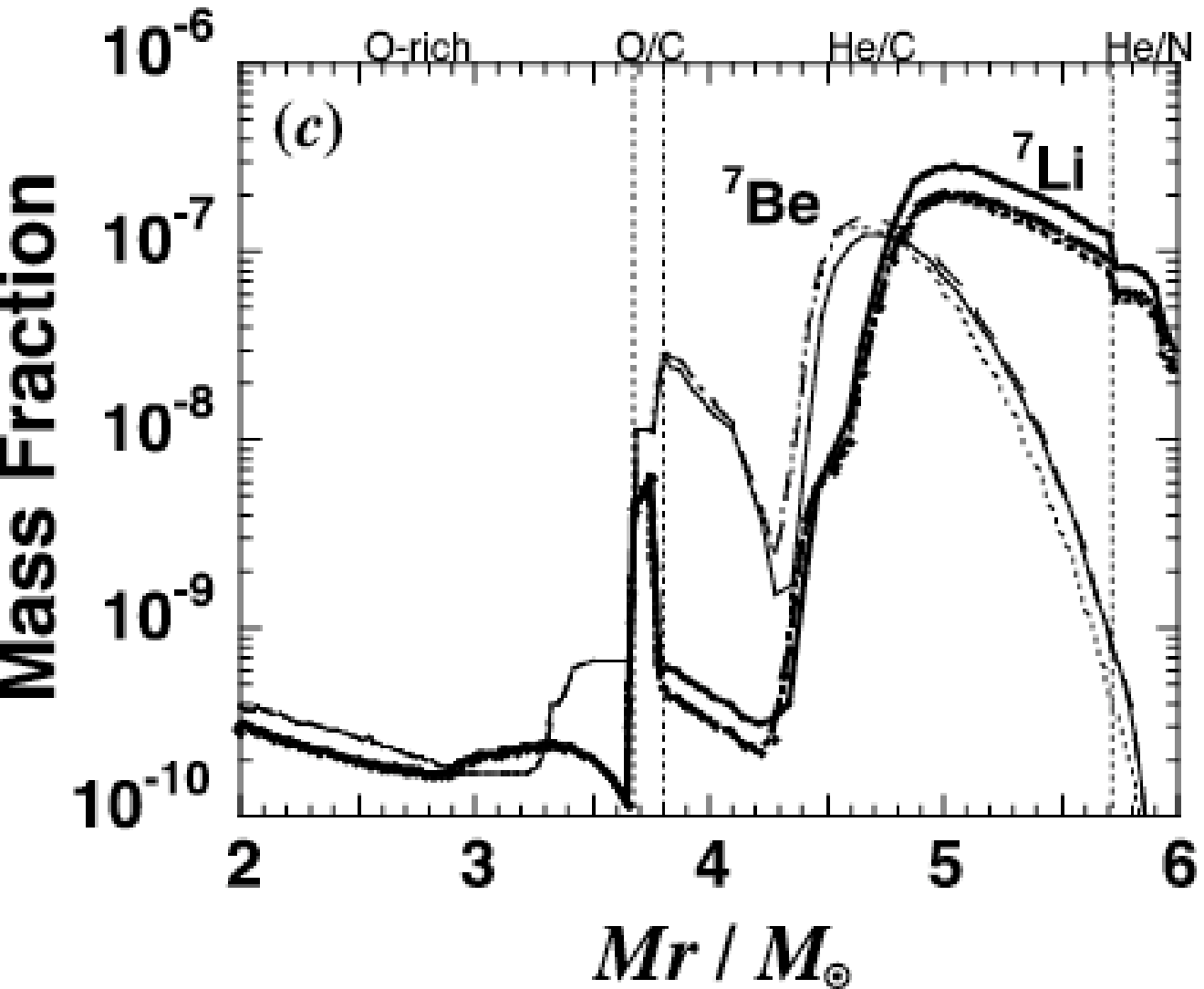}
\plotone{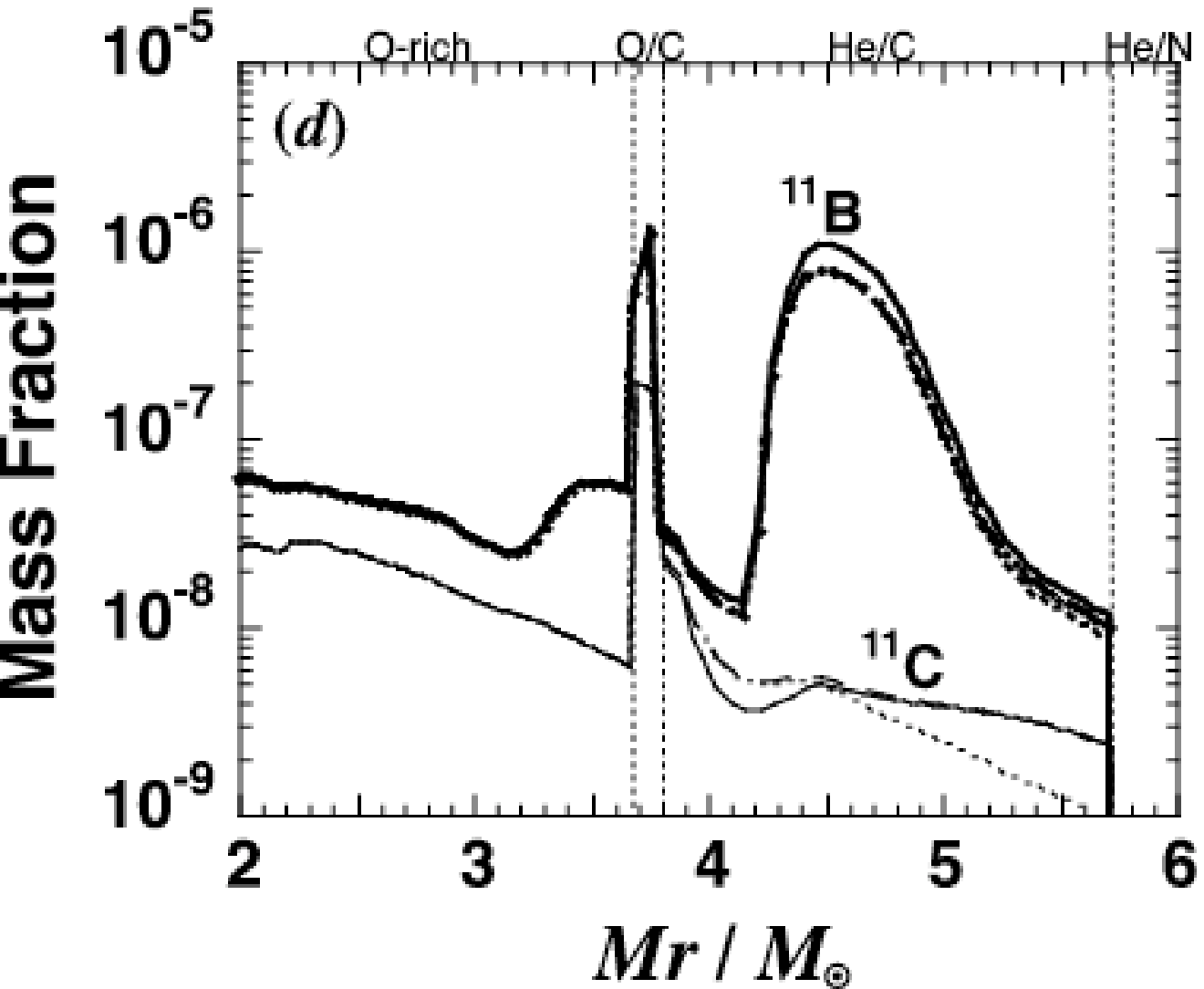}
\caption{
{\it Continued}.
}
\end{figure}

\begin{figure}
\epsscale{0.7}
\plotone{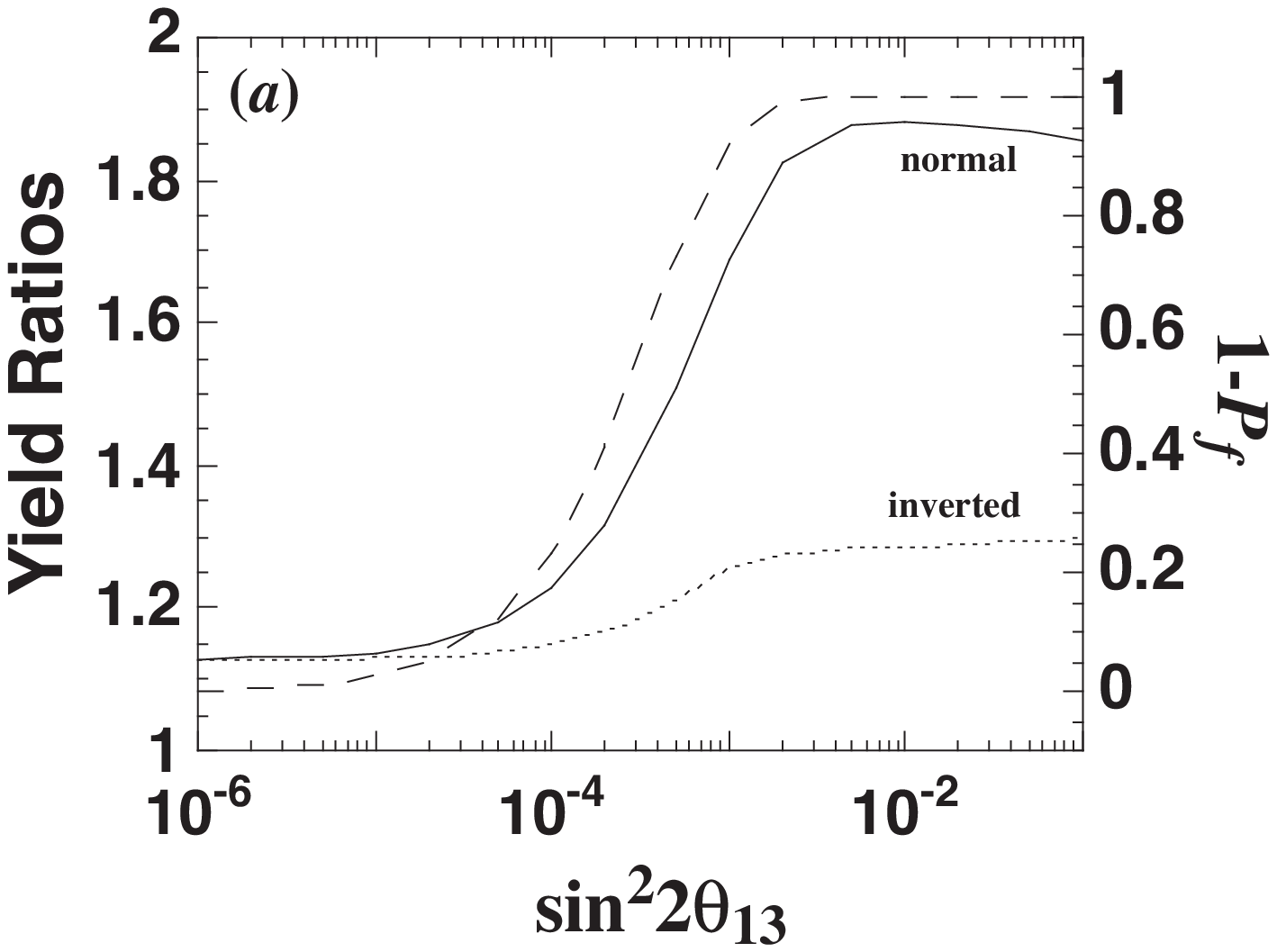}
\epsscale{0.6}
\plotone{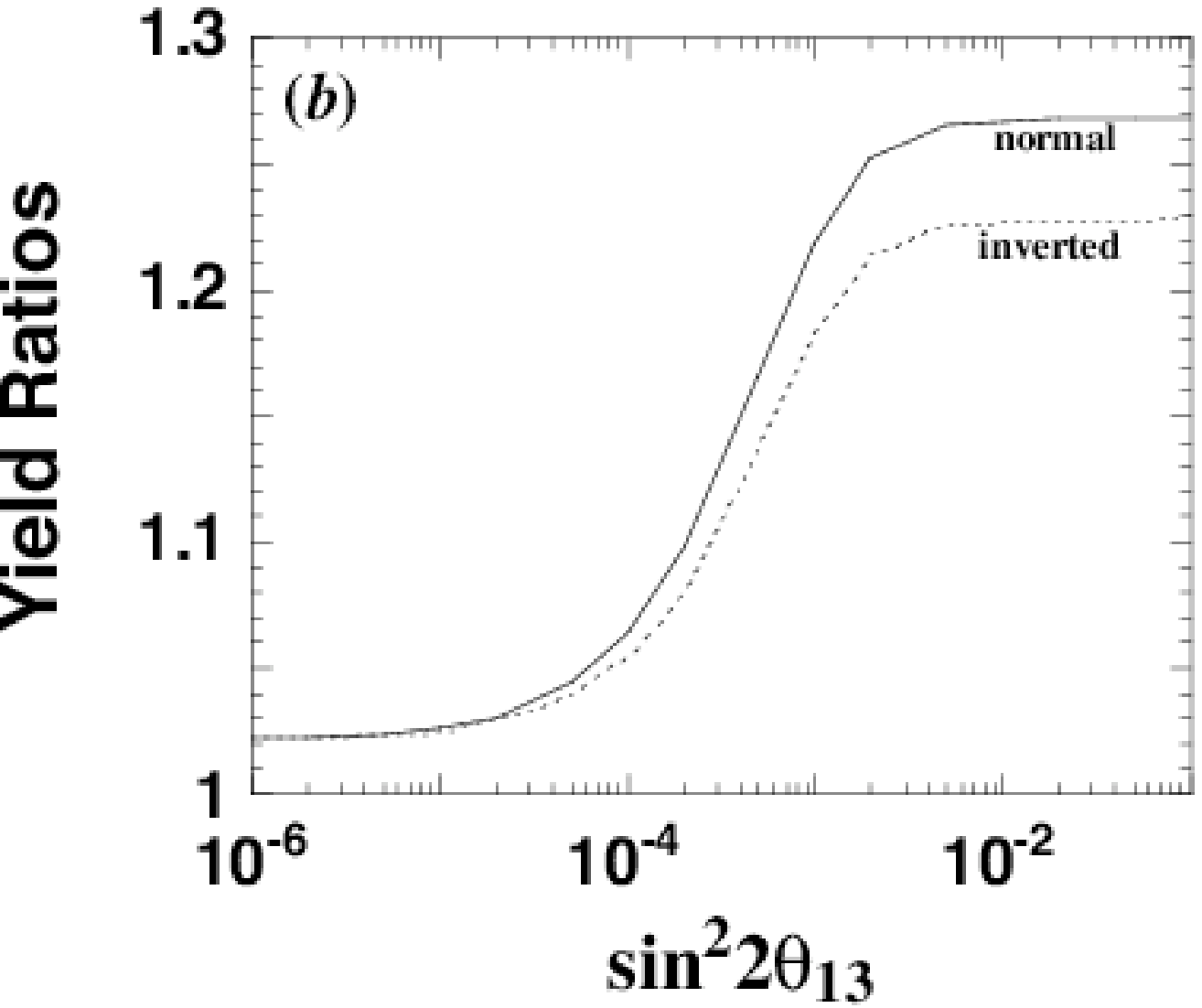}
\caption{
Dependence of the yield ratios of ({\it a}) $^7$Li and ({\it b}) $^{11}$B 
on $\sin^{2}2\theta_{13}$.
Solid line and dotted line indicate the yield ratios in the normal and 
inverted mass hierarchies.
In panel ({\it a}), dashed line indicates the flip probability in the case of 
neutrino energy equal to 50 MeV as the formulation of $1-P_f$.
}
\end{figure}


\clearpage
\begin{figure}
\epsscale{0.6}
\plotone{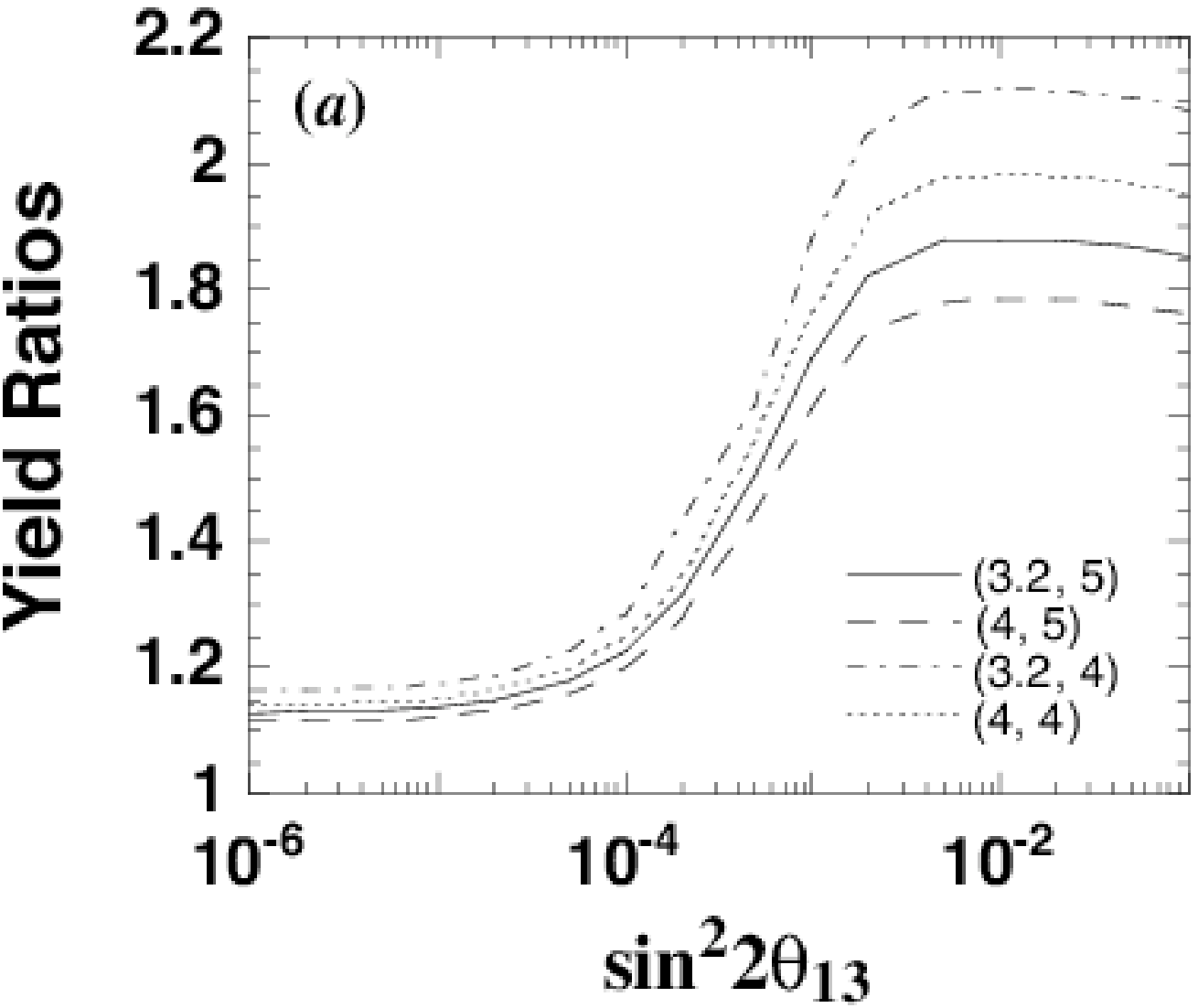}
\plotone{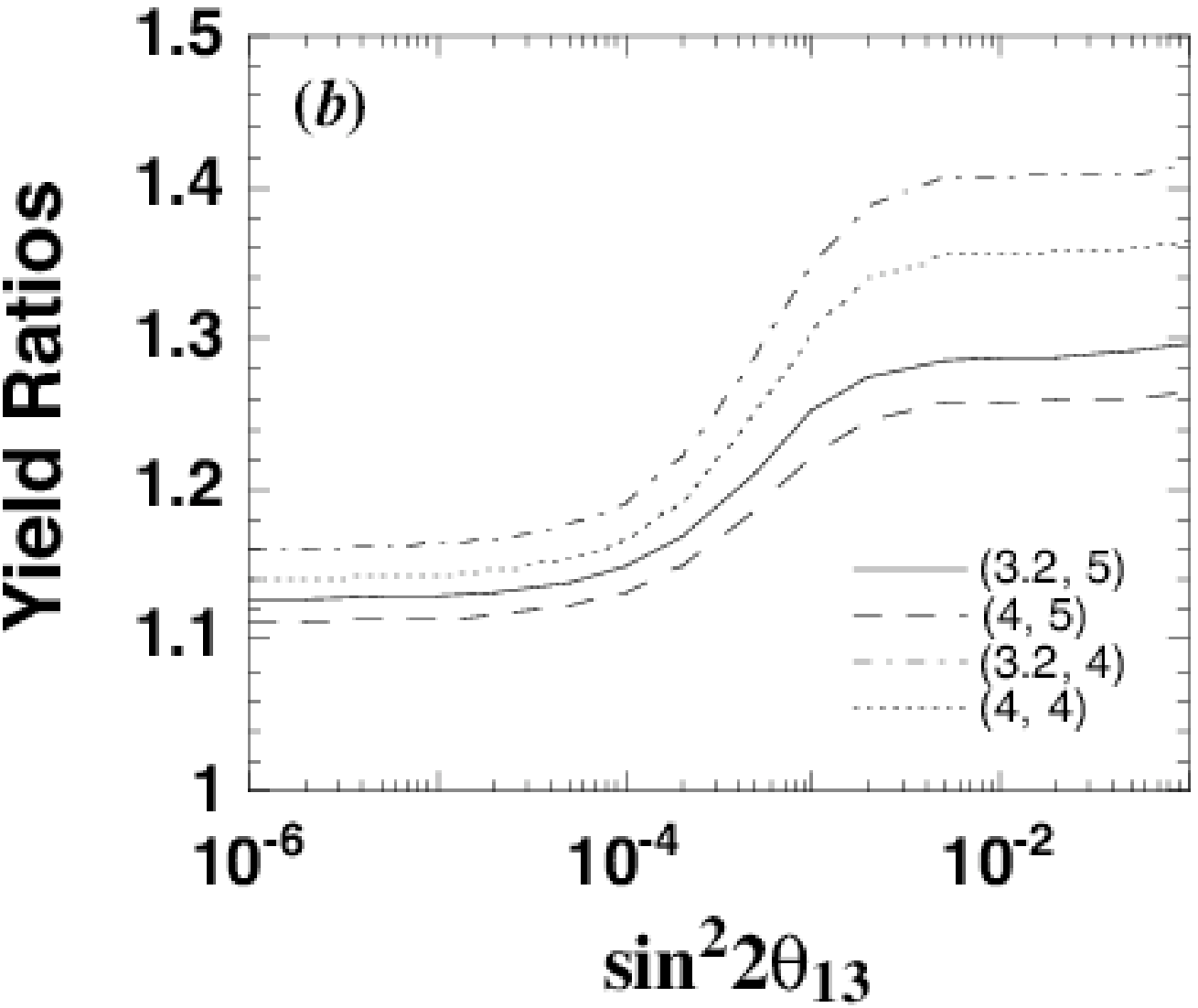}
\caption{
Yield ratios of $^7$Li (({\it a}) and ({\it b})) and $^{11}$B 
(({\it c}) and ({\it d})) related to the temperatures of $\nu_e$ and 
$\bar{\nu}_e$ emitted from a proton-neutron star.
Panels ({\it a}) and ({\it c}) correspond to the normal mass hierarchy and 
panels ({\it b}) and ({\it d}) correspond to the inverted mass hierarchy.
Solid, dashed, dash-dotted, and dotted lines indicate the yield ratios
in the cases of ($T_{\nu_e},T_{\bar{\nu}_e}$) = 
(3.2 MeV, 5 MeV), (4 MeV, 5 MeV), (3.2 MeV, 4 MeV), and (4 MeV, 4 Mev).
}
\end{figure}

\begin{figure}
\figurenum{5}
\plotone{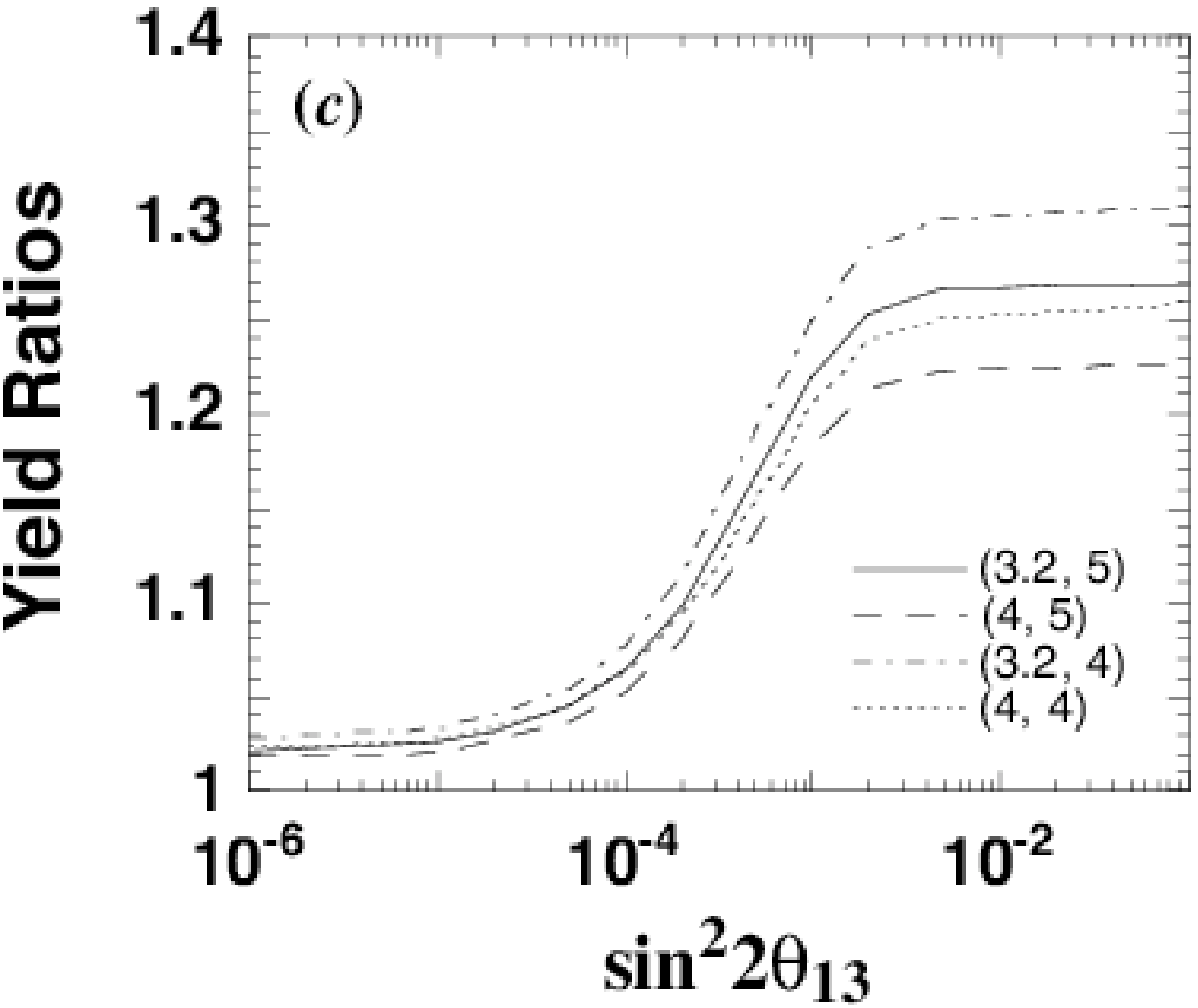}
\plotone{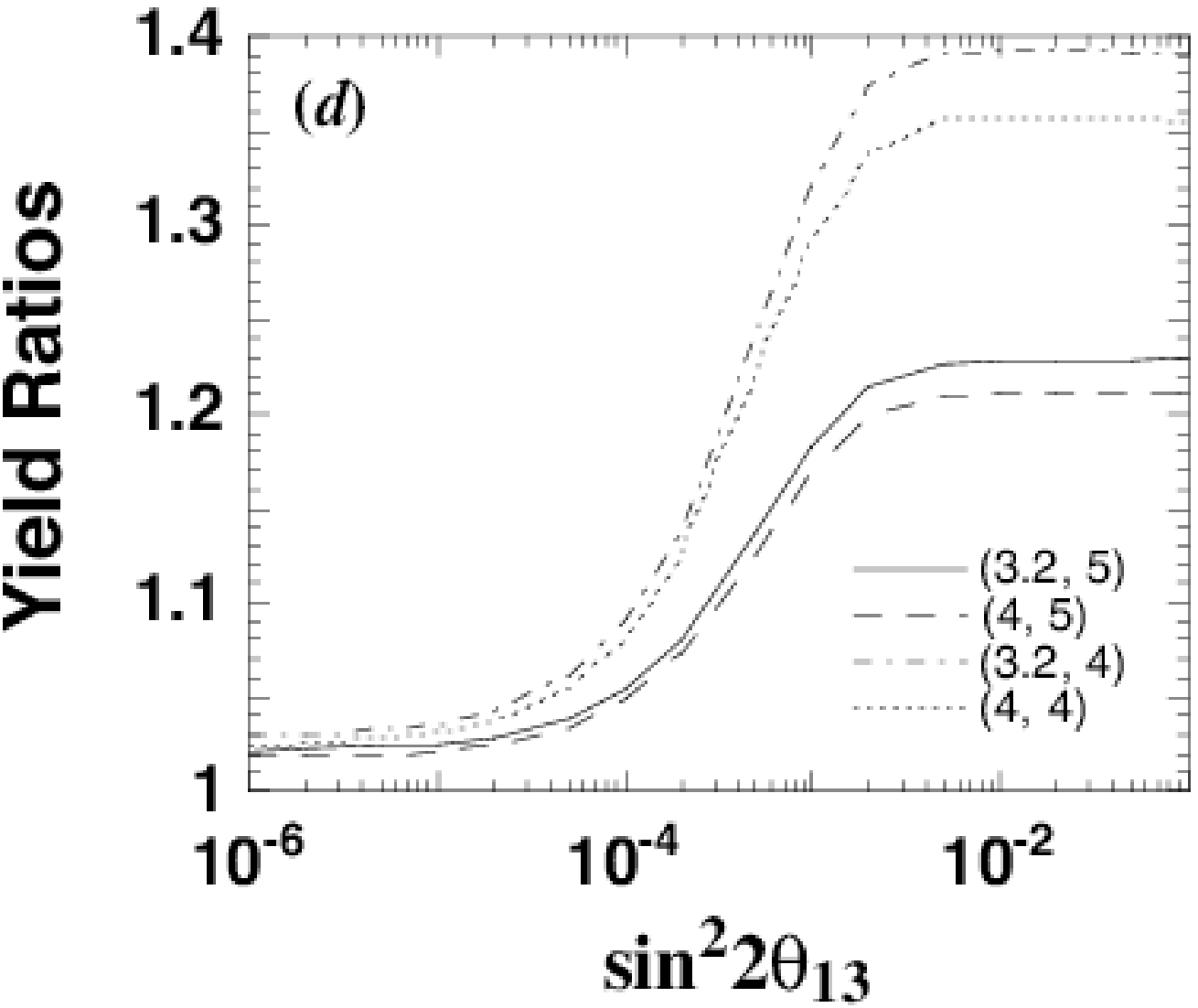}
\caption{
{\it Continued}.
}
\end{figure}

\clearpage
\begin{figure}
\plotone{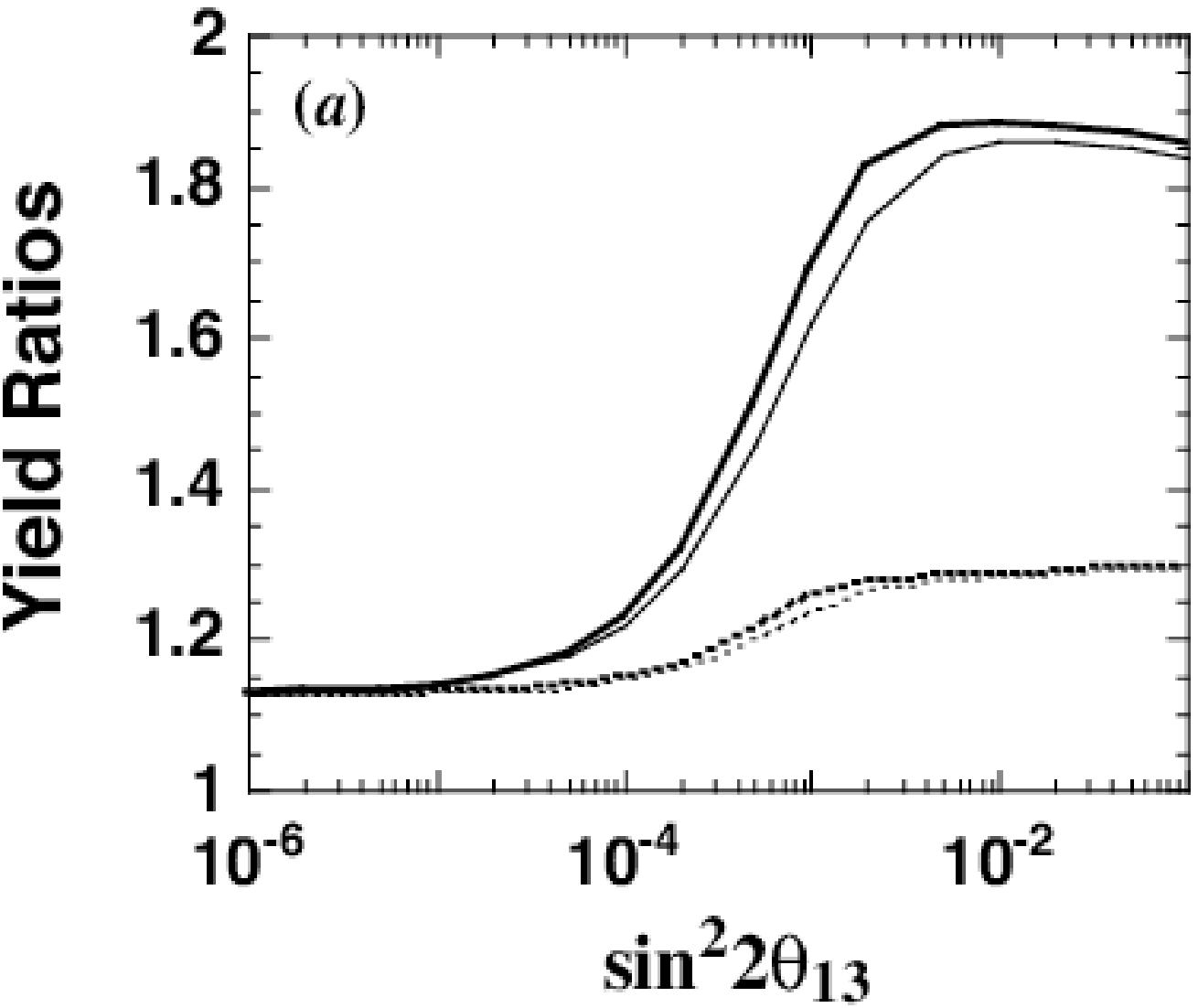}
\plotone{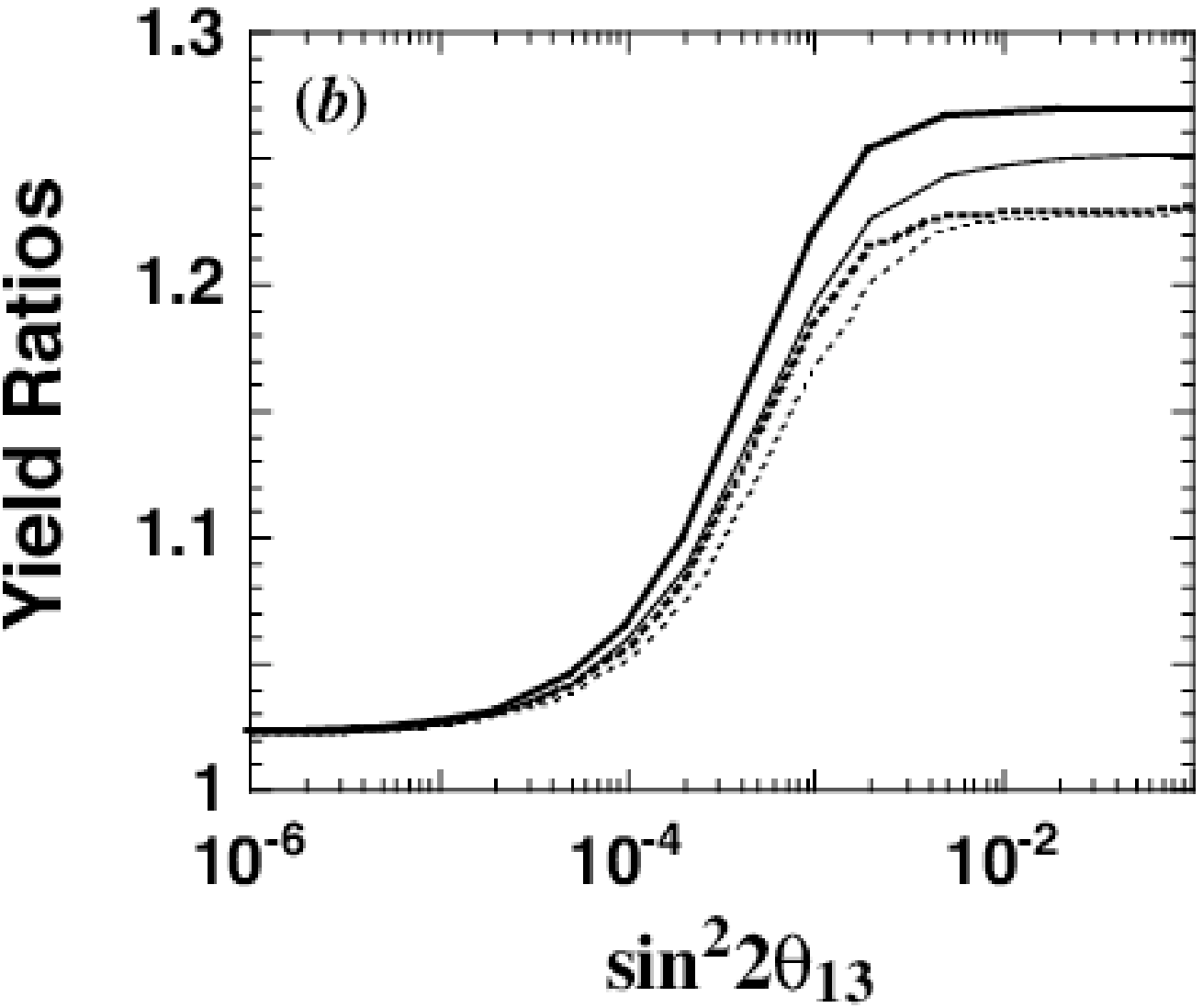}
\caption{
Dependence of the ({\it a}) $^7$Li and ({\it b}) $^{11}$B yield ratios on 
$\sin^{2}2\theta_{13}$ taking account of the shock propagation effect on
neutrino oscillations.
Solid lines and dotted lines correspond to the normal and inverted mass 
hierarchies.
Thick lines are the same as in Fig. 4.
Thin lines indicate the yield ratios taking into account the shock 
propagation effect.
}
\end{figure}

\clearpage
\begin{figure}
\plotone{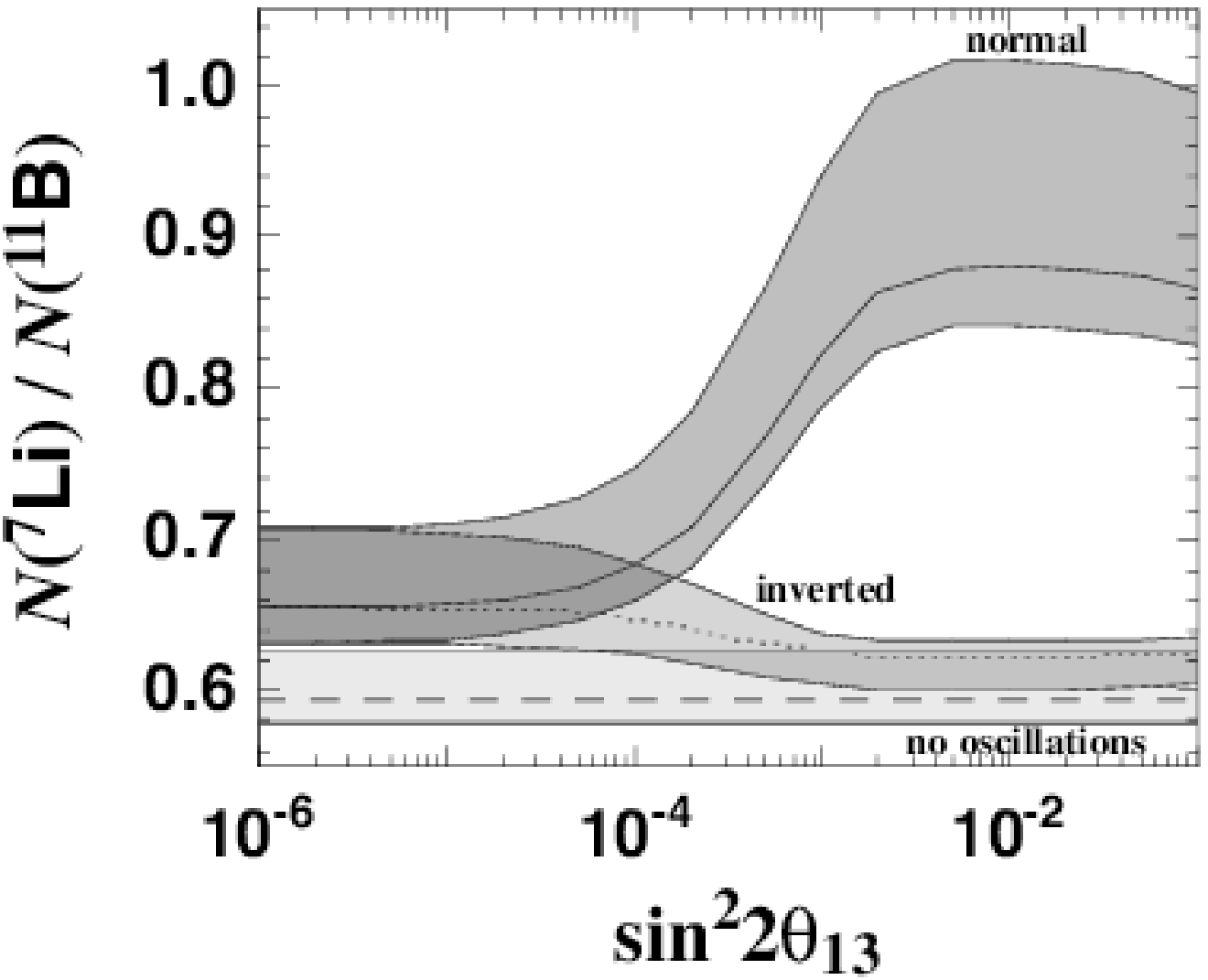}
\caption{
Dependence of the $^7$Li/$^{11}$B number ratio on $\sin^{2}2\theta_{13}$.
The dark- and medium-shaded regions indicate the ranges of the 
$^7$Li/$^{11}$B ratios in the normal and inverted mass hierarchies.
The number ratio without the neutrino oscillations is indicated by the
light-shaded region.
These regions are obtained from the evaluations using four sets of the 
temperatures of $\nu_e$ and $\bar{\nu}_e$ defined in \S 4.3.
Solid, dotted, and dashed lines in each region correspond to the number
ratio with our standard model of the neutrino temperatures (see \S 4.3).
See text for more details.
}
\end{figure}






\clearpage

\begin{table}
\begin{center}
\caption{Coefficients and threshold energies of the cross sections
for charged-current reactions on $^4$He and $^{12}$C.
The evaluated cross sections are in units of $10^{-42}$ cm$^2$.
}
\vspace{5mm}
\begin{tabular}{cccc}
\tableline\tableline
Reaction & $a$ & $b$ & $\varepsilon_{th}$ (MeV) \\
\tableline
$^4$He($\nu_e,e^-p)^3$He & 
$2.45027 \times 10^{-6}$ & 4.03496 & 19.795 \\
$^4$He($\bar{\nu}_e,e^+n)^3$H & 
$2.37357 \times 10^{-5}$ & 3.31394 & 21.618 \\
$^{12}$C($\nu_e,e^-p)^{11}$C & 
$3.73589 \times 10^{-6}$ & 3.78171 & 17.939 \\
$^{12}$C($\bar{\nu}_e,e^+n)^{11}$B & 
$2.76414 \times 10^{-5}$ & 3.08738 & 17.761 \\
$^{12}$C($\nu_e,e^-\gamma)^{12}$N & 
$2.45749 \times 10^{-3}$ & 1.88899 & 17.3381 \\
$^{12}$C($\bar{\nu}_e,e^+\gamma)^{12}$B & 
$2.13890 \times 10^{-3}$ & 1.72299 & 14.3909 \\
\tableline
\end{tabular}
\end{center}
\end{table}

\begin{table}
\begin{center}
\caption{Yields of $^7$Li and $^{11}$B without neutrino oscillations
for four sets of the adopted temperatures of $\nu_e$ and $\bar{\nu}_e$.
The temperature of $\nu_{\mu,\tau}$ and $\bar{\nu}_{\mu,\tau}$ is assumed
commonly to be 6.0 MeV (see \S 2.2).
}

\begin{tabular}{ccc}
\tableline\tableline
($T_{\nu_e},T_{\bar{\nu}_e}$) & $M({\rm ^7Li})/M_\odot$ & 
$M({\rm ^{11}B})/M_\odot$ \\
\tableline
(3.2 MeV, 5 MeV)  & $2.36 \times 10^{-6}$ & $6.26 \times 10^{-6}$  \\
(4 MeV, 4 MeV)    & $2.24 \times 10^{-6}$ & $5.77 \times 10^{-6}$  \\
(4 MeV, 5 MeV)    & $2.50 \times 10^{-6}$ & $6.80 \times 10^{-6}$  \\
(3.2 MeV, 4 MeV)  & $2.08 \times 10^{-6}$ & $5.23 \times 10^{-6}$  \\
\tableline
\end{tabular}
\end{center}
\end{table}




\end{document}